\documentclass[twocolumn]{aastex631}

\usepackage{apjfonts}
\DeclareSymbolFont{cmletters}{OML}{cmm}{m}{it}
\DeclareMathSymbol{v}{\mathalpha}{cmletters}{"76}

\usepackage{graphicx}	
\usepackage{amsmath}	
\usepackage{amssymb}	
\usepackage{amsfonts,amsthm,epsfig,float,tabularx}
\usepackage{wrapfig}
\usepackage{soul}
\usepackage{placeins}
\usepackage{xcolor}
\usepackage{hyperref}

\usepackage{comment}

\newcommand{\appropto}{\mathrel{\vcenter{
  \offinterlineskip\halign{\hfil$##$\cr
    \propto\cr\noalign{\kern2pt}\sim\cr\noalign{\kern-2pt}}}}}

\shorttitle{Effect of Gas Angular Momentum on the Jet Launching}
\shortauthors{Kwan et al.}
\graphicspath{{./}{figures/}}

\begin{document}

\title{The Effects of Gas Angular Momentum on the Formation of Magnetically Arrested Disks and the Launching of Powerful Jets}

\correspondingauthor{Lixin Dai lixindai@hku.hk}

\author[0000-0003-0509-2541]{Tom M. Kwan}
\affiliation{Department of Physics, The University of Hong Kong, Pokfulam Road, Hong Kong}

\author[0000-0002-9589-5235]{Lixin Dai}
\affiliation{Department of Physics, The University of Hong Kong, Pokfulam Road, Hong Kong}

\author[0000-0002-9182-2047]{Alexander Tchekhovskoy}
\affiliation{Department of Physics \& Astronomy, Northwestern University, Evanston, IL 60208, USA}
\affiliation{Center for Interdisciplinary Exploration \& Research in Astrophysics (CIERA), Evanston, IL 60208, USA}



\begin{abstract}

 In this letter, we investigate Bondi-like accretion flows with zero or low specific angular momentum by performing 3D general relativistic magnetohydrodynamic simulations. In order to check if relativistic jets can be launched magnetically from such flows, we insert a large-scale poloidal magnetic field into the accretion flow and consider a rapidly spinning black hole.
We demonstrate that under such conditions the accretion flow needs to initially have specific angular momentum above a certain threshold to eventually reach and robustly sustain the magnetically arrested disk state. If the flow can reach such a state, it can launch very powerful jets at $\gtrsim 100\%$ energy efficiency. Interestingly, we also find that even when the accretion flow has initial specific angular momentum below the threshold, it can still launch episodic jets with an average energy efficiency of $\sim 10\%$. However, the accretion flow has nontypical behaviors such as having different rotation directions at different inclinations and exhibiting persistent outflows along the midplane even in the inner disk region.  
Our results give plausible explanations as to why jets can be produced from various astrophysical systems that likely lack large gas specific angular momenta, such as Sgr A*, wind-fed X-ray binaries, tidal disruption events, and long-duration gamma-ray bursts.

\end{abstract}



\section{Introduction} \label{sec:intro}

\noindent Gas accretion onto black holes (BHs) powers many luminous astrophysical systems, such as active galactic nuclei (AGNs), BH X-ray binaries, and tidal disruption events (TDEs).
The accreting gas usually forms a disk or torus around the BH, in which the magnetorotational instability (MRI)  \citep{Balbus.1998} can transport the gas angular momentum outward and allow the gas to accrete. Accretion converts the mass-energy of the gas into the energy of relativistic jets, winds and radiation produced from the accreting system.

Some of the popular accretion disk models are the advection-dominated accretion flow \citep[ADAF; e.g.,][]{Narayan.1994.ADAF,Narayan.1995.ADAF, Yuan.2014}, standard $\alpha$-disk \citep{Novikov.1973.NT, Shakura.1973.standard.disk} and slim disk \citep{Abramowicz.1988} models, which apply to disks at different accretion rates. While all these disk models assume the disk has a relatively large specific angular momentum, some astrophysical systems can carry very low angular momentum gas, such as Sgr A* \citep{Melia.1992, Narayan.1995, Ressler.2018}, long gamma-ray bursts \citep[GRBs;][]{Fryer.1999}, TDEs \citep{Rees.1988} and wind-fed high-mass X-ray binaries \citep[HMXBs;][]{Smith.2002,Tauris.2006}. Such low angular momentum accretion flows are less widely studied, yet it is important to understand whether and how they can produce powerful emission, winds, and jets.

A spherical symmetric steady-state accretion flow (with zero angular momentum and no magnetic field) is described by the Bondi accretion model \citep{Bondi.1952} under Newtonian physics. 
The behavior of Bondi-like accretion flows with low angular momentum has also been extensively studied with hydrodynamic simulations on large scales \citep[e.g.][]{Proga.2003, Proga.2003b, Lee.2006, Janiuk.2008, Sukova.2015, Sukova.2017, Murguia.2020}, which show that a thick torus structure can form as the flow hits the centrifugal barrier and part of the inflow becomes an outflow as a result.
However, these simulations have not included general relativity or the magnetic field, either of which is expected to play an essential role in determining the behavior of inner accretion flows and, in particular, their prospect of launching relativistic jets \citep{Blandford.1977}. 

The dynamical evolution of magnetized plasma around BHs and the production of jets involve nonlinear processes and generally lack obvious symmetries. This makes general relativistic magnetohydrodynamic (GRMHD) simulations particularly attractive for modeling these systems \citep[e.g.,][]{DeVilliers.2003, Gammie.2003, Tchekhovskoy.2011, McKinney.2012, White.2016, Porth.2017, Porth.2019}. 
These GRMHD simulations usually start from gas in the configuration of a torus or disk with its motion-dominated rotation \citep[see reviews by][]{Tchekhovskoy.2012a, Tchekhovskoy.2015, Davis.2020}. They show that if a large amount of ordered magnetic field is supplied, the accretion flow drags sufficient magnetic flux to the vicinity of the BH, which will form magnetically arrested disks (MADs; \citealt{Narayan.2003}; see also \citealt{Bisnovatyi-Kogan.1974,Bisnovatyi-Kogan.1976}). In MADs, the magnetic field significantly impacts the inner disk dynamics, and  
very powerful jets can be produced if the BH spins rapidly \citep{Blandford.1977}. In fact, the jet power can even exceed the accretion power, resulting in net energy extraction from the BH \citep{Tchekhovskoy.2011}.

In contrast, the behavior of low angular momentum accretion flows has not been extensively explored with GRMHD simulations. 
\citet{Komissarov.2009} studied the spherical accretion of magnetized gas with 2D GRMHD simulations and showed that such accretion flow can become highly magnetized near the event horizon and launch (nonrelativistic) outflows.
More recently, several studies employed 3D GRMHD simulations to investigate Bondi-like accretion flow and the resulting feedback processes. \citet{Lalakos.2022} connected the BH to its sphere of influence and found that only a small fraction ($\sim 1 \%$) of the gas reaches the BH when the ambient medium has a low angular momentum. \citet{Ressler.2021} studied spherical accretion flow threaded with a large-scale magnetic field at different inclinations with respect to the spin axis of the BH and explored how the jet strength depends on the inclination. \citet{Nicholas.2022} simulated Bondi-Hoyle-Lyttleton (BHL) accretion onto a spinning BH traversing a magnetized gaseous medium and showed such accretion could power relativistic jets. However, how the behavior of the accretion flow and the power of the jet depend on the gas angular momentum has not been systematically studied.

In this work, we focus on investigating the role that the gas specific angular momentum plays in Bondi-like accretion flow. We carry out three fully 3D GRMHD simulations of magnetized Bondi-like accretion flows with initially zero or very low angular momentum around rapidly spinning BHs. 
We study the properties of the quasi-steady-state accretion flow, and investigate how the magnetic flux and gas coexist and interact near the BH. We demonstrate that both nonrotating and slowly rotating accretion flows can produce relativistic jets. However, the initial angular momentum of the gas needs to reach a small critical value to robustly sustain the MAD state and its powerful jets.
We introduce our methodology and simulation setup in Section \ref{sec:method}. We present the results in Section \ref{sec:results}, and summarize and discuss them in Section \ref{sec:discussion}.

\begin{deluxetable*}{cccccccccccccc}
\label{tab:parameter}
\tablecaption{Parameter Summary of the simulations \label{tab:summary}}
\tablewidth{0pt}
\tablehead{
\colhead{Model} &   \colhead{$R_c$ ($r_g$)} &   \colhead{$N_r$} &   \colhead{$N_{\theta}$} &   \colhead{$N_{\phi}$} & 
 \colhead{$\phi_{H}$} & 
 \colhead{$\eta_{H} (\%)$} & \colhead{$\eta_{\rm{}jet} (\%)$} & \colhead{$\eta^{\rm{}EM}_{\rm{}jet} (\%)$} & \colhead{$s$}
}
\startdata
a09rc0	&   0   &   288 &   128 &   64    &	26.8	&	 19.9	&	 17.1	&	   13.0	 &	-12.0  \\
a09rc10	&   10  &   240 &   128 &   64   &	23.2	&	   10.8	&	 9.8	&	   6.8	&	-10.5	 \\
a09rc50	&   50  &   280 &   128 &   64   &	45.1	&	 76.1	&	 61.7	&	   52.6	&	-35.2	
\enddata
\end{deluxetable*}

\section{Methodology} \label{sec:method}

\subsection{Governing Equations}

We carry out 3D time-dependent GRMHD simulations using the HARM code \citep{Gammie.2003,McKinney.2004,Tchekhovskoy.2011,McKinney.2012}. 
The code solves the following conservation equations:
\begin{align}
    \nabla_{\mu} \left(\rho u^{\mu}\right) &= 0 \\
    \nabla_{\mu} T^{\mu}_{\nu} &= 0
\end{align}
where $\rho$ is the rest-mass density of the gas in the fluid frame, $u^{\mu}$ is the gas four-velocity, and $T^{\mu}_{\nu}$ is the stress--energy tensor, which includes matter (MA) and electromagnetic (EM) terms:
\begin{equation} \label{eq1}
    \begin{split}
    T^{\mu}_{\nu} &=  {T^{\rm EM}}^{\mu}_{\nu} + {T^{\rm MA}}^{\mu}_{\nu}, \\
	{T^{\rm MA}}^{\mu}_{\nu} &= (\rho + u_g + p_g) \ u^{\mu}u_{\nu}+ p_g \delta^{\mu}_{\nu},  \\
	{T^{\rm EM}}^{\mu}_{\nu} &= b^2 u^{\mu}u_{\nu} + p_b \delta^{\mu}_{\nu}-b^{\mu}b_{\nu},  
	\end{split}
\end{equation} 
where $u_g$ is the gas internal energy density with an adiabatic index $\gamma=4/3$, $p_g=(\gamma-1)u_{g}$ is the gas pressure;  $p_b=b^{\mu}b_{\nu}/2=b^2/2$ is the magnetic pressure; $b^{\mu}$ and $b_{\nu}$ are the contravariant and covariant fluid-frame magnetic field four-vectors, respectively; and $\delta^{\mu}_{\nu}$ is the Kronecker delta tensor. Here and below we use Lorentz-Heaviside units. The magnetic field is evolved using the induction equation:
\begin{equation} \label{eq:magnetic_field}
    \partial_t (\sqrt{-g} B^i) = -\partial_j \left(\sqrt{-g} (b^j u^i - b^i u^j) \right)
\end{equation}
where $g$ is the metric determinant and $B^i$ is the magnetic field three-vector.

\subsection{Simulation Setup and Initial Conditions} \label{sec:setup}

We consider a rapidly rotating BH with a dimensionless spin parameter $a=0.9$ and conduct simulations for three different initial specific gas angular momenta. The simulations employ modified spherical polar coordinates centered on the BH, with a minimum resolution of $N_r \times N_{\theta} \times N_{\phi} = 240\times128\times64$ cells in the r-, $\theta$-, and $\phi$-direction (see Table \ref{tab:summary} for more details). The grid in the $r$-direction is logarithmically spaced. The grid in the $\theta$-direction has higher resolution in the disk (equatorial) and jet (polar) regions in order to resolve the turbulence in the equatorial disk and the collimated polar jets. The grid in the $\phi$-direction is uniform.
The simulation domain extends from the inner radial boundary at $R_{\rm{}in} = 0.8 r_H \approx 1.15 r_g$ to the outer radial boundary at $R_{\rm{}out}=10^5 r_g$. This ensures that there are at least five grid cells between $R_{\rm in}$ and $r_H$, thereby ensuring that the inner radial boundary is causally disconnected from the accretion system. Here, $r_g=GM_{\rm{}BH}/c^2$ is the BH gravitational radius, where $G$ is the  gravitational constant, $c$ is the speed of light, $M_{\rm{}BH}$ is the mass of the BH, and $r_H = (1+ \sqrt{1-a^2}) r_g$ is the BH event horizon radius.
Nonradiative ideal GRMHD simulations are scale-free in BH mass, so we use geometric units with $G = c = M_{\rm{}BH}= 1$ hereafter. 

For standard Bondi accretion, a spherically symmetric gas distribution falls inward, attracted by the gravity of the central BH. At infinity, gas elements have specific energy $e=1$, which is conserved during the infall. Therefore, an approximate self-similar solution for the gas density $\rho$ and four-velocity $u^\mu$ is

\begin{align} 
    \rho & \propto r^{-3/2},  \label{eq:density} \\
    u_t & = - e = -1, \label{eq:ut} \\
    u^\theta &= 0, \label{eq:utheta}\\
    u_\phi  &= 0, \label{eq:uphi} \\
    u^r &= - \frac{\sqrt{2r (r^2+a^2)}}{r^2+a^2\cos^2\theta}. \label{eq:ur} 
\end{align}

For the simulated gas flow with zero or low angular momentum, while we still use Equation (\ref{eq:density})--(\ref{eq:utheta}) to initialize $\rho$, $u_t$, and $u^\theta$, we modify the initial $u_\phi$ and $u^r$ according to the desired initial specific angular momentum.
The specific angular momentum of a Keplerian equatorial circular orbit is
\begin{equation}
    l_K (r) = \frac{\sqrt{r}(r^2-2a\sqrt{r}+a^2)}{r\sqrt{r^2-3r+2a\sqrt{r}}} .
    \label{eq:l_K}
\end{equation}
We adopt the initial specific angular momentum characterized by the circularization radius $R_c$, at which the centrifugal force is strong enough to counterbalance the inward gravitational force. That is, our gas in the equatorial plane at $r>R_{\rm solid}=100 r_g$ initially has a constant specific angular momentum corresponding to $R_c = 0$ (model a09rc0), $10 r_g$ (model a09rc10), or $50 r_g$ (model a09rc50): 

\begin{equation}
    l_{\rm{}max} =
    \begin{cases}
    0                   & \text{for} \ R_c=0 \\
    l_K (r= R_c)        & \text{for} \ R_c = 10 r_g  \ {\rm or} \  50 r_g \\
    \end{cases}
    \label{eq: angular_momentum}
\end{equation}
The specific angular momentum at $r<R_{\rm solid}$ or $\theta \neq \pi/2$ is set not to exceed $l_{\rm max}$ (see Appendix \ref{sec:initial} for details). The gas is also isentropic with $u_g/\rho^{\gamma} = \rm{constant}$.

We thread the accretion flow with a relatively weak poloidal magnetic field. The magnetic flux is described by the magnetic flux function $\Psi (r,\theta) = r (1-|\cos\theta|)$, which describes parabolic magnetic flux distribution, for which $B^2 \propto r^{-2}$ asymptotically far away from the BH \citep{Tchekhovskoy.2008}. Since $u_g \propto \rho^{\gamma} \propto (r^{-3/2})^{\gamma} \propto r^{-2}$ in our initial conditions, this magnetic flux ensures that the plasma beta $\beta\equiv p_g / p_b$ is asymptotically constant everywhere. We normalize the magnetic field by rescaling it such that $\beta$ has a minimum value of 100 along the equator, so that the initial magnetic field is subdominant.

\subsection{Diagnostics}
\label{sec:diag}

We divide the gas flow into three regions: 1) the jet region, where $\sigma= b^2/\rho >1$;  
2) the wind region, where $u^r>0$ and $\sigma<1$; and 3) the disk region, where $u^r<0$ and $\sigma<1$. (We note that a small inflow region near the BH can have $\sigma >1$, but excluding this region from the disk does not affect our results.\footnote[4]{In practice, here we measure $\dot{M}_H$, $\dot{E}_H$ and $\eta_H$ at $r=6r_g$ to avoid potential contamination due to the numerical density floors near the horizon, since the true time-averaged $\dot{M}_H$ and $\dot{E}_H$ (if uncontaminated by the density floors) should remain constant near the horizon when the flow has reached a quasi-steady state.})
The above conditions define the boundaries between these regions, i.e., the jet--wind boundary has $\sigma=1$ and the disk--wind boundary has $u^r=0$. 
Below we summarize the definitions we use for computing various quantities. 

The rate of rest-mass accretion flowing inward through a sphere of radius $r$ is
\begin{equation}
	\dot{M} (r)= -\int \rho u^r \text{d} A_{\theta\phi} , \label{eq:mdot}
\end{equation} where $dA_{\rm\theta\phi}=\sqrt{-g} \rm d\theta \rm d\phi$. This gives us the rate of accretion onto the BH $\dot{M}_H$ when it is evaluated at the event horizon.\textsuperscript{4}

The absolute magnetic flux through the BH event horizon is
\begin{equation}
    \Phi_H = \frac{1}{2} \int |B^{r}|  \,  \text{d} A_{\theta\phi} \label{eq:Phi}
\end{equation}
where the integral is over the event horizon. 
We define a dimensionless magnetic flux parameter, the ``MADness parameter,'' by normalizing $\Phi_{H}$ with $\dot{M}_H$ \citep{Tchekhovskoy.2011}:
\begin{equation}
\phi_{H}=\frac{\Phi_H}{\sqrt{\dot{M}_H r^{2}_{g} \ c}} 
\label{eq:phi}
\end{equation}
where for clarity we restore the dimensional factors and use the CGS units.
The disk enters the MAD state when $\phi_{H}\gtrsim40$ \citep{Tchekhovskoy.2011, Tchekhovskoy.2015}.
In practice, when we compute dimensionless quantities like $\phi_H$ and the below efficiencies by normalizing some quantity $Q(t)$ with $\dot{M}_H (t)$, we smooth $\dot{M}_H (t)$ with a Gaussian function over a moving window centered at time $t$ and with a standard deviation of $1000 ~r_g/c$ in order to reduce fast fluctuations.

The energy flux flowing out through a sphere of radius $r$ is
\begin{equation}
    \dot{E} (r) = -\int T^r_t  \text{d} A_{\rm\theta\phi} .
\end{equation} 
Using this, we can calculate the outflow efficiency as
\begin{equation}
	\eta (r) \equiv \frac{\dot{E}(r)+\dot{M}(r)}{\dot{M}_H} .
	\label{eq:eta_net}
\end{equation}
The accretion efficiency $\eta_H$ is defined as $\eta (r)$ evaluated at the event horizon. When $\eta_H > 100\%$, accretion results in net energy extraction out of the BH, dominated by the extraction of energy in electromagnetic form through the Blandford-Znajek process out of a rapidly spinning BH, so that the extracted energy exceeds the accretion-supplied energy in the form of rest-mass energy. In addition, we can evaluate the jet energy flux and the jet energy efficiency as
\begin{equation}
    \dot{E}_{\rm{}jet} (r) = -\int T^r_t(\sigma>1) \text{d} A_{\rm\theta\phi},
\end{equation}
and
\begin{equation}
	\eta_{\rm jet} (r) \equiv \frac{\dot{E}_{\rm jet}(r)}{\dot{M}_H},
	\label{eq:eta_jet}
\end{equation}
respectively.

In order to obtain the radial profile of a quantity $Q(r)$, we average it over spherical shells and weight it by the gas density $\rho$:
\begin{equation}
    \langle Q(r) \rangle_{\rho}\equiv \frac{\int  \text{d} A_{\rm \theta\phi}\rho \textit{Q}}{\int  \text{d} A_{\rm \theta\phi}\rho} ,
\end{equation}
where the integral is over a sphere of radius $r$. We can then define the geometric half angular thickness, $H/R$, of the flow (after excluding the jet region) as
\begin{equation}
    \frac{H}{R}\equiv \left( \frac{\int \rho (\theta - \theta_{\rm{}mid})^2 (\sigma<1)\; {\rm d}A_{\rm \theta \phi}}{\int \rho (\sigma<1) \; {\rm d}A_{\theta\phi}}\right) ^{1/2} ,
\end{equation}
where $\theta_{\rm{}mid} (r, \phi) \equiv \int \rho \theta \ \text{d} \theta /\int \rho \ \text{d} \theta$ is the polar angle of the disk midplane at a specific ($r$, $\phi$) location.

Most quantities will be computed and plotted in geometric units throughout the paper. We provide a table to translate the units to CGS units in Appendix \ref{sec:convert}.


\section{Results}
\label{sec:results}

\subsection{Time Evolution of the Accretion Flow }
\label{sec:results_time_evolution}

\begin{figure*}
  %
  \begin{center}
    \includegraphics[width=0.85\textwidth]{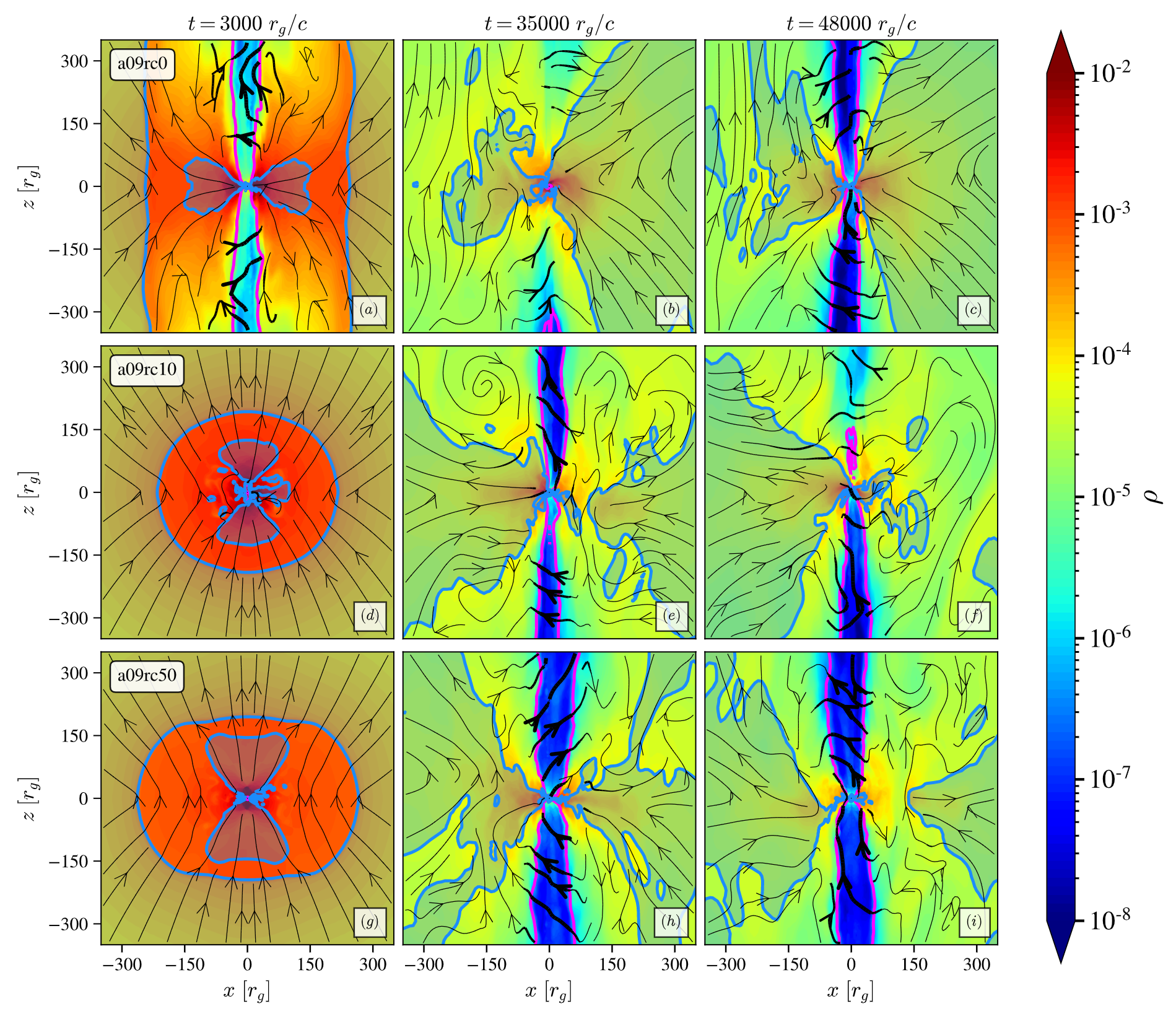}
    \vspace*{-0.5cm}
  \end{center}
\caption[]{We show snapshots of a vertical slice from the inner regions of the 3D GRMHD simulations with a09rc0 ($R_c=0$, top), a09rc10 ($R_c=10 r_g$, middle), or a09rc50 ($R_c=50 r_g$, bottom), at different times $t=$ (3000, 35,000, 48,000) $r_g/c$ (left to right). The colors show the gas rest mass density $\rho_{0}$. The black lines indicate the magnetic field lines, of which the thickness represents the magnetic field strength. The pink contours show the boundary of the jet with $\sigma=1$. The blue contours show the disk--wind (i.e., inflow--outflow) boundary, which has $u^r=0$. In all simulations, geometrically thick disk structures (shaded with a light blue color) form close to the BHs. In addition, wide-angle winds and relativistic jets are launched. 
}
\label{fig:structure}
\end{figure*} 

We run all three simulations out to at least $t = 50,000 r_g/c$ and show their evolution in Figure \ref{fig:structure}. The flows reach an inflow equilibrium out to $r \gtrsim 100 r_g$ in all runs.
The three runs differ in their value of $R_c$.
Initially, the gas falls in radially from large distances without sufficient centrifugal support. A centrifugally supported structure soon starts to form close to the BH, owing to two factors: 1) A small initial gas specific angular momentum (as in our models a09rc10 and a09rc50) can result in a centrifugal barrier, which has been seen in analytical calculations and hydrodynamical simulations of Bondi-like accretion \citep[e.g.,][]{Proga.2003, Lee.2006, Sukova.2017}. 2) Due to the general relativistic frame-dragging effect, the gas near the BH can gain some angular momentum. Therefore, even in our a09rc0 model, the spherical symmetry is broken due to the spin of the BH and the vertical configuration of the magnetic flux.  
After the formation of the centrifugal barrier, the infalling gas hits it and develops a shock, which propagates outward. A high-density bubble expands with the shock (e.g., the regions confined by two blue $u^r =0$ contours in the snapshots at $t=3000 r_g/c$). 
By $t \gtrsim 10,000 r_g/c$, the accretion flows have become stable in all three simulations, and winds and relativistic jets are launched. The jets drive blast waves into the ambient medium, push out the gas, and prevent accretion onto the BH along the polar region.

\begin{figure*}
    \includegraphics[width=\textwidth]{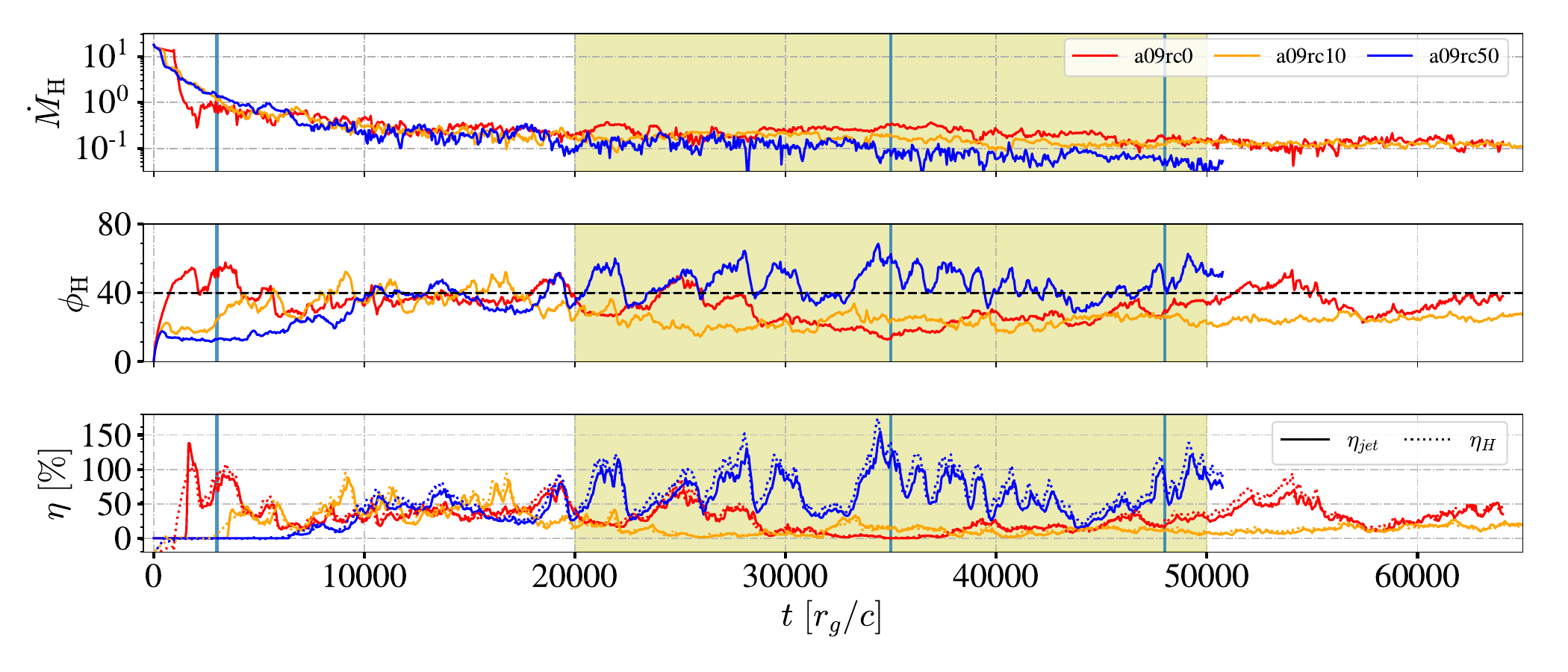}
    \vspace*{-0.8cm}
    \caption[caption with time evolution]{Time evolution of the mass accretion rate $\dot{M}_H$ (top panel), dimensionless magnetic flux $\phi_H$ (middle panel), outflow energy efficiency $\eta_H$ (bottom panel, dotted line), and jet efficiency $\eta_{\rm{}jet}$ (bottom panel, solid line) of the three models a09rc0 (red), a09rc10 (orange), and a09rc50 (blue). The jet efficiency is measured at $r=100 r_g$; the other quantities are measured close to the horizon. The three vertical blue lines indicate the time at which the snapshots in Figure \ref{fig:structure} are taken. The light yellow highlight indicates the time interval over which we compute the time-averaged values of the various quantities. The horizontal dashed line in the middle panel indicates the critical value of $\phi_H \sim 40$ for the disk to enter the MAD state. 
    The smaller the value of $R_c$ and hence of the initial specific angular momentum, the faster the magnetic fluxes accumulate initially. However, the MAD state and very powerful jet are only sustained in the model a09rc50 with the highest initial specific angular momentum.  }
    \label{fig:time_evolution}
\end{figure*}

Figure \ref{fig:time_evolution} shows the evolution of the rate of mass accretion onto the BH $\dot{M}_H$ (Equation (\ref{eq:mdot})), the MADness parameter $\phi_{H}$ (Equation (\ref{eq:phi})), the outflow energy efficiency $\eta_{H}$ (Equation (\ref{eq:eta_net})) and the jet efficiency $\eta_{\rm jet}$ (Equation (\ref{eq:eta_jet})) as functions of time. 
The rate at which gas falls into the BH starts out high. Dragged by the infalling gas, magnetic flux builds up near the BH, so $\phi_{H}$ quickly grows, and MRI is triggered in the accretion disk. It is apparent that the model a09rc0 has the $\phi_{H}$ that grows the fastest, since the gas in this model falls in the fastest with the lowest initial gas angular momentum. After the disk forms, $\dot{M}_H$ drops with time, until a later phase at $t\sim20,000 r_g/c$ when the accretion becomes quasi-steady.

Interestingly, though magnetic fluxes accumulate faster in models a09rc0 and a09rc10 in the initial phase, after $t \sim 20,000 r_g/c$, $\phi_{H}$ in these two models decreases and eventually reaches a value below the MAD threshold. In contrast, we see $\phi_{H}$ grows the slowest in model a09rc50, which has the largest initial specific angular momentum. However, it can reach and sustain the MAD state and demonstrate characteristic MAD behavior: high variability in $\Phi_{H}$ and $\eta$ and the production of powerful jets with $\eta_{\rm jet}>100\%$ from the Blandford--Znajek process ($\eta_{\rm jet}$ is plotted using a solid line in the bottom panel). In addition, there is a significant correlation between $\eta_{H}$ and $\eta_{\rm jet}$, meaning most of the energy produced in the accretion process is carried away by jets.

 We will perform more analysis of the behavior of magnetic flux in Section \ref{sec:dissipation}. Furthermore, see Appendix \ref{sec:mri} for calculations related to the MRI, where we show that the MRI is resolved in all three models and is mildly suppressed in models a09rc10 and a09rc50 but not in a09rc0.

\subsection{The Radial and Angular Structure of the Flow}
\label{sec:results_radial&inclination}

\begin{figure*}
    \centering
    \includegraphics[width=0.65\textwidth]{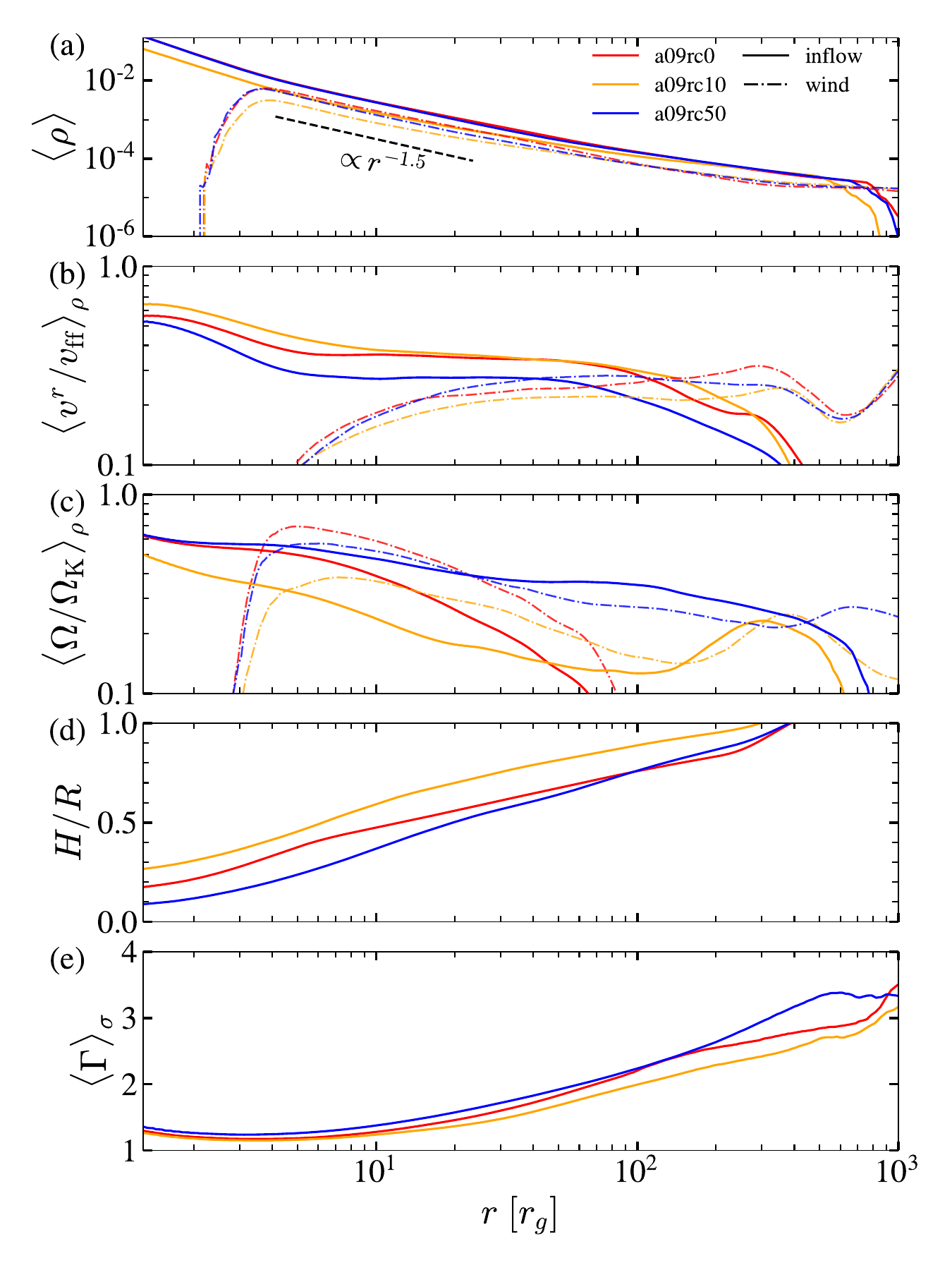}
    \vspace*{-0.5cm}
    \caption{The time and angle-integrated radial profiles of (a) density $\rho$, (b) lab-frame radial velocity $v^r$ ($\rho$-weighted) over the freefall velocity $v_{\rm ff}$, (c) lab-frame angular velocity $\Omega$ ($\rho$-weighted) over the relativistic Keplerian angular velocity $\Omega_{\rm K}$, (d) geometric half angular disk thickness $H/R$, and (e) Lorentz factor of the jet $\Gamma$ ($\sigma$-weighted). The three different colors denote the different models: a09rc0 (red), a09rc010 (yellow), and a09rc50 (blue). The quantities in the inflow and wind regions in (a)-(c) are plotted using solid and dotted--dashed lines, respectively. All models produce sub-Keplerian, sub-freefall, and geometrically thick disks, though the model a09r50 produces the thinnest and fastest-rotating disk among the three. The relativistic jets in the models have a peak Lorentz factor $\Gamma=3-4$, at $r\sim 1000 r_g$.
  }
    
    \label{fig:radial_profile}
\end{figure*}

\noindent In this section we investigate the radial and inclination profiles of accretion flow in the window of $t=20,000$--$50,000 \ r_g/c$. In this time window, both gas accretion and $\phi_{H}$ have become quasi-steady in the inner regions of the accretion flow.

Figure \ref{fig:radial_profile} shows the radial profiles of (a) the gas density, $\rho$; (b) the ratio of the lab-frame radial velocity, $v^r=u^r/u^t$, to the freefall velocity $v_{\rm ff}$ (in practice $v^r/v_{\rm ff} = -u^r / u^r_{\rm ff}$ with $u^r_{\rm ff}$ given by Equation (\ref{eq:ur})); (c) the ratio of the lab-frame angular velocity, $\Omega=u^\phi/u^t$, to the Keplerian velocity, $\Omega_{\rm K}=1/(r_p^{1.5}+a)$, where $r_p=r\sin\theta$ is the cylindrical radius; (d) the geometric half angular thickness of the inflow, $H/R$; and (e) the Lorentz factor, $\Gamma = \alpha u^t$, where $\alpha=1/\sqrt{-g^{tt}}$ is the time lapse.
All quantities are time-$\theta$-$\phi$-averaged. In addition, the quantities in (b)--(c) are density-weighted to focus on the disk, while $\Gamma$ is magnetization-weighted to focus on the jet. 
We plot these quantities out to $1000 r_g$, where the shocks have propagated to, so the $\rho$, $v^r$ and $\Omega$ of the inflows decline rapidly around that radius.

One can see that the gas density and radial velocity structures of the accretion flows in the three models are similar in the $r\lesssim 100 r_g$ inflow equilibrium region. 
All models produce geometrically thick disks, while model a09rc50 (with the largest specific angular momentum among the three) produces the least thick disk. The disks completely extend to the event horizon, and their densities still approximately follow the freefall profile $\rho \propto r^{-1.5}$. The radial inflow velocities are 30-40\% of the freefall velocity in the quasi-steady inflow region. In particular, a09rc50 has lower $v^r$ than the other two models. Moreover, despite the different initial conditions, all three accretion inflows acquire some angular momentum and rotate at the frequency of a few $\times 0.1 \Omega_{\rm{K}}$ near the horizon. However, only the model a09rc50 can maintain this near-Keplerian rotation until $r \sim 700 r_g$, while the rotation speed in the other two models quickly declines at larger radii. 
Furthermore, in all models winds are launched close to the BH and rapidly accelerated mostly due to the high magnetic pressure. They reach the maximum density within $r = 5 r_g$ and the terminal velocity within $r = 100 r_g$. Lastly, the jets launched in the three models have similar acceleration profiles and eventually reach $\Gamma= 3-4$. The jet in the model a09rc50 is steadier and slightly faster. 

\begin{figure*}[ht]
\centering
\includegraphics[width=0.65\textwidth]{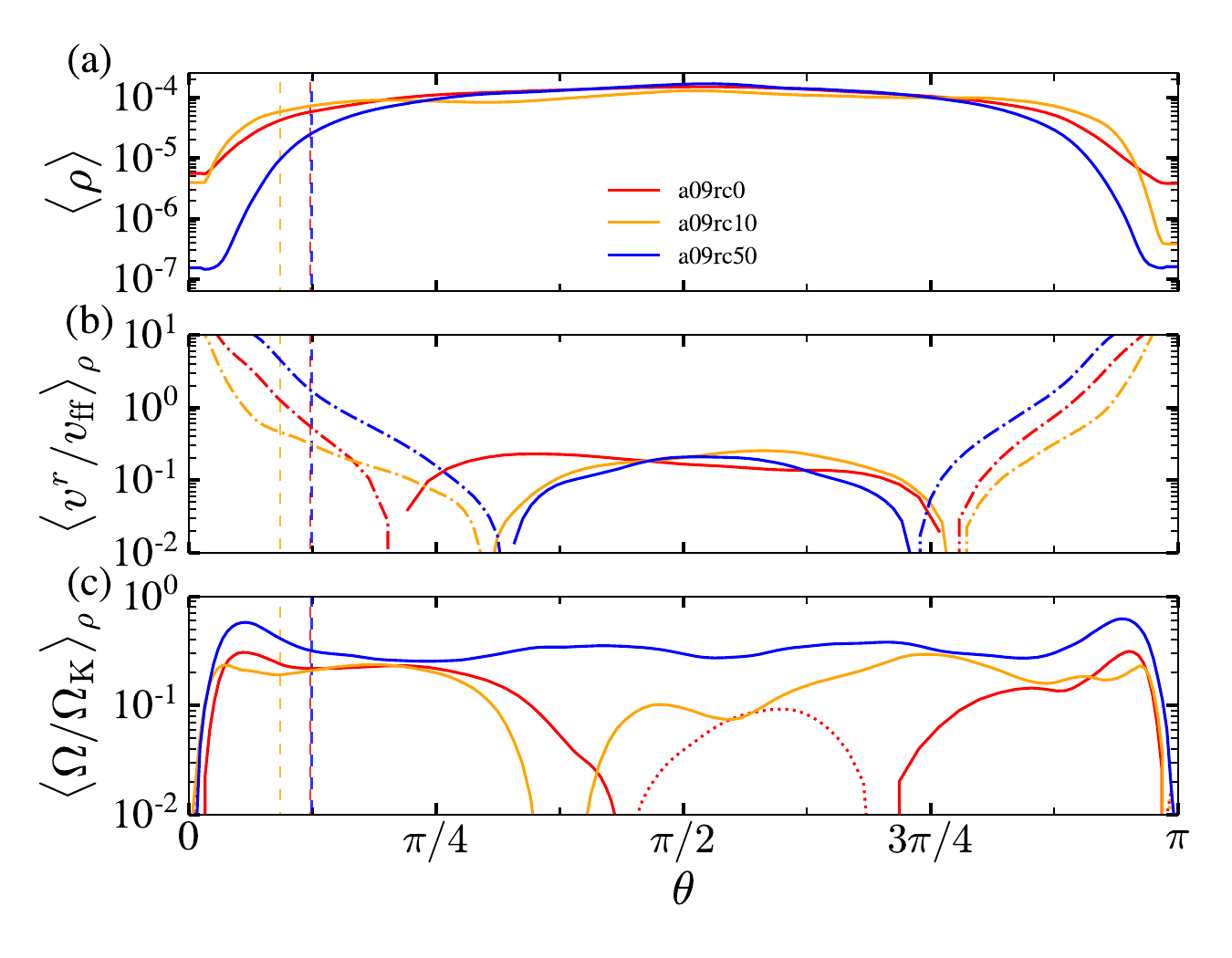}
\caption{The time and $\phi$-averaged $\theta$-profiles of (a) $\rho$, (b) $v^r/v_{\rm ff}$ ($\rho$-weighted), and (c) $\Omega/\Omega_{\rm K}$ ($\rho$-weighted) for the surface at $r=100 r_g$. The colors indicate the three models as in Figure \ref{fig:radial_profile}. In panel (b), solid lines indicate inflow, and dotted-dashed lines indicate outflow. In panel (c), solid lines indicate clockwise rotation, and dotted lines indicate counterclockwise rotation. The vertical dashed lines indicate the northern jet boundary with $\sigma=1$ in the three models. Only the accretion flow in model a09rc50 can develop coherent rotation among the three. Model a09rc50 also produces the highest radial and angular speed of the wind.
} 
\label{fig:angular_profile}
\end{figure*}

Figure \ref{fig:angular_profile} shows the $\theta$-profiles of (a) $\rho$,  (b) $v^r/ v_{\rm ff}$ and (c) $\Omega/\Omega_{\rm K}$, all evaluated on the surface at $r=100r_g$. These quantities are time/$\phi$ averaged. Consistent with the $H/R$ result, the density profile of the disk in model a09rc50 is the most concentrated toward the midplane. Indeed the a09rc50 disk most closely resembles the disks produced in typical GRMHD simulations starting from torus setups, though geometrically thicker. The $v^r$ plot also clearly shows the inflow--outflow boundary moves toward the midplane as the gas specific angular momentum increases. The winds reach maximum radial velocities of $v_r \approx 0.1 c$ right outside the jet in models a09rc0 and a09rc10, while in the a09rc50 model the wind reaches a higher velocity of $v_r \approx 0.2 c$. The wind in a09rc50 also has the highest angular velocity of $\Omega \approx$ 0.3--0.4$\Omega_{\rm K}$, while the winds in models a09rc0 and a09rc10 only rotate at $\Omega \approx$ 0.1--0.2$\Omega_{\rm K}$. 
Most interestingly, the behavior of the angular velocity $\Omega$ in model a09rc50 and that in the other two models are drastically different. Only in model a09rc50 does the accretion inflow develop a coherent rotation at all inclinations, while the gas in a09rc0 and a09rc10 still orbits in different directions at different inclinations, due to the conservation of angular momentum.  

Lastly, the radial and angular profiles of the magnetic fields are also calculated. We refer the reader to Appendix \ref{sec:Magnetic_field} for more details.

\subsection{Behavior of Magnetic Flux and Transport of Angular Momentum in the Accretion Flow}
\label{sec:dissipation}
In the final part of the study, we focus on how the magnetic flux accumulates (or leaks out) and helps transport angular momentum in these Bondi-like accretion flows. 

\begin{figure*}
\vspace*{-1.7cm}
    \begin{tabular}[b]{@{}p{0.1\textwidth}@{}}
        \vspace*{4.0cm}
        \gridline{\fig{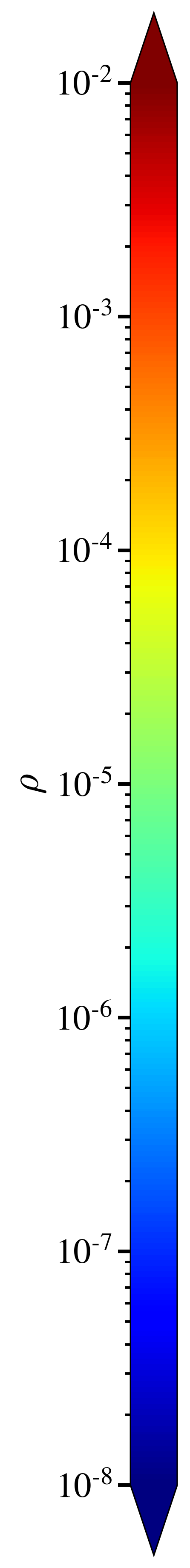}{0.1\textwidth}{} }
    \end{tabular}
    \hspace*{-1.0cm}
    \begin{tabular}[b]{@{}p{0.75\textwidth}@{}}
        \gridline{\fig{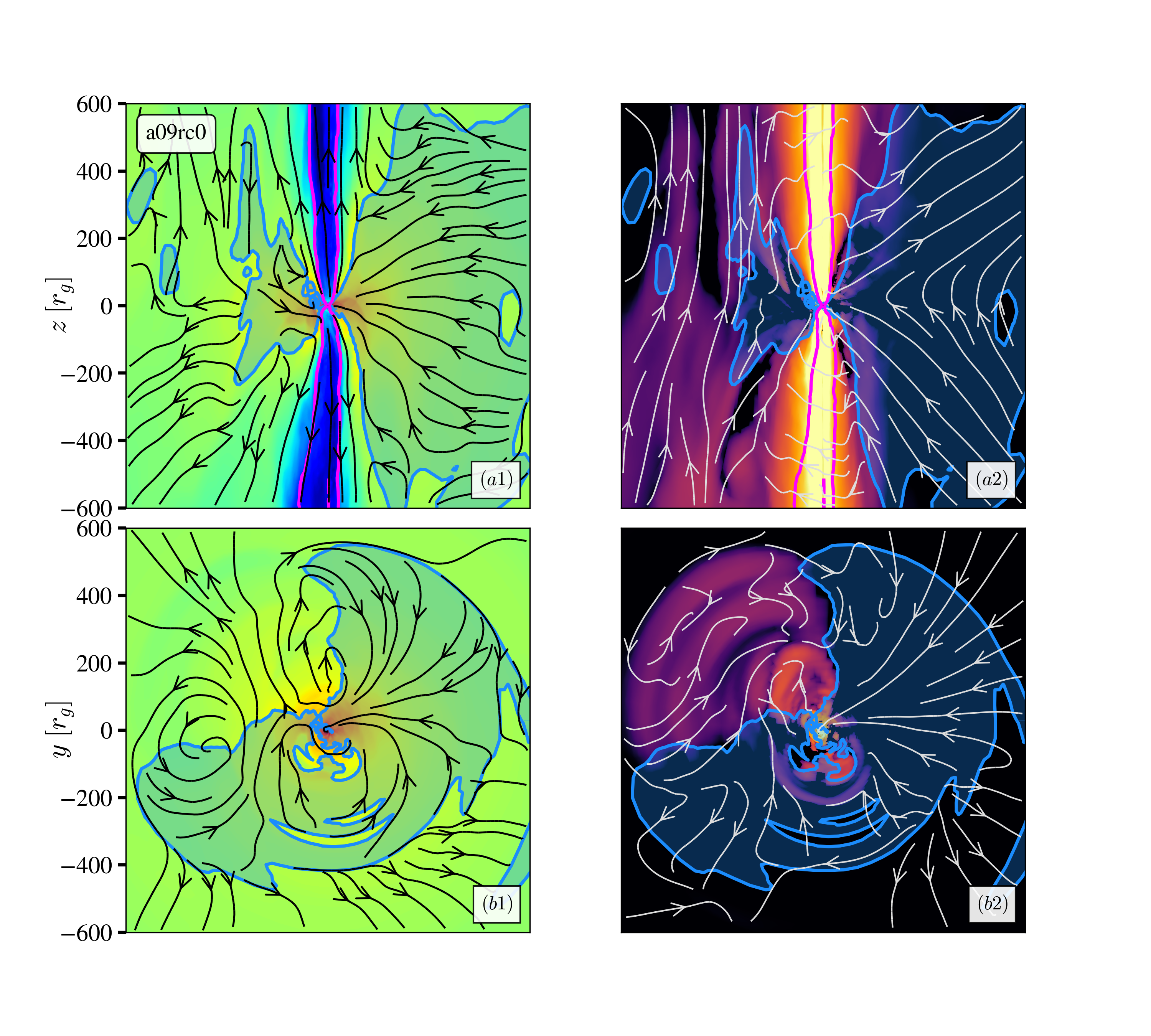}{0.75\textwidth}{} \label{fig:leakage_rc0}}
        \vspace*{-3.0cm}
        \gridline{\fig{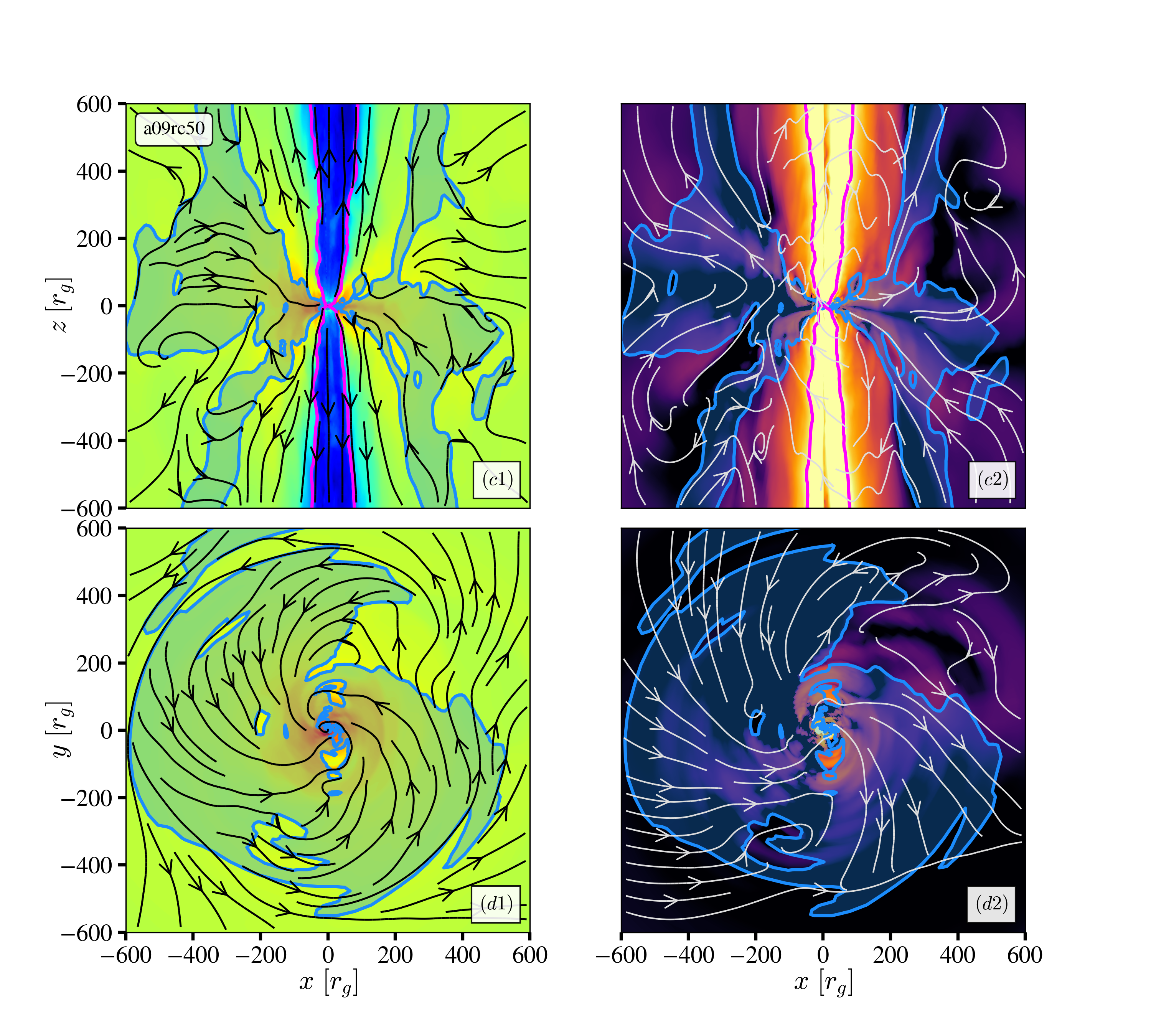}{0.75\textwidth}{} \label{fig:leakage_rc50}}
    \end{tabular}%
    \hspace*{-2.0cm}
    \begin{tabular}[b]{@{}p{0.108\textwidth}@{}}
        \vspace*{4.0cm}
        \gridline{\fig{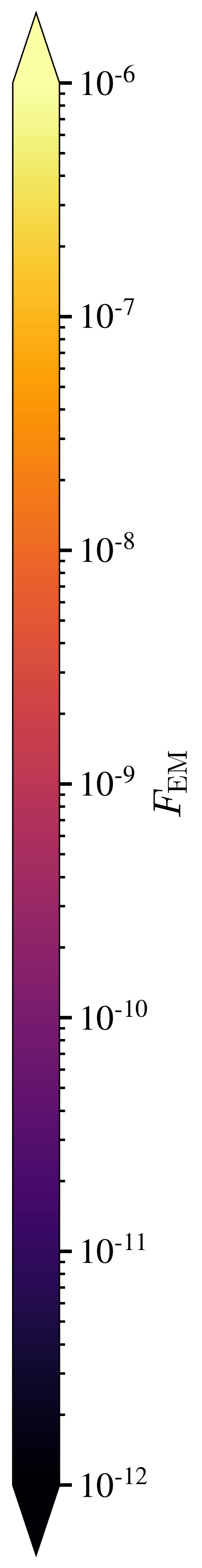}{0.107\textwidth}{} }
    \end{tabular}
    
    \vspace*{-1.3cm}
\caption{\textbf{Leakage of magnetic flux:} The panels show snapshots at $t=35,000 r_g/c$ in the a09rc0 model (top four panels) and a09rc50 model (bottom four panels) in the vertical plane (panels (a) and (c)) and equatorial plane (panels (b) and (d)).  The left column shows the gas density, and the right column shows the electromagnetic energy flux.  The inflow regions are shaded with a light blue color, with blue contours marking the inflow--outflow boundaries. The pink contours indicate the jet boundaries with $\sigma=1$. The black lines in the left column are the density-weighted velocity streamlines, and the white lines in the right column are the magnetic field lines. One can see that in the a09rc0 model gas around the midplane can persistently flow out at a large range of angles and carry away large magnetic fluxes. In contrast, in the $R_c=50 r_g$ run, a steadier inflow structure forms, which likely plays an important role in retaining the high-level magnetic flux in the inner accretion flow. (Full movies of these models can be seen at \url{https://tomkm.com/madbondi.html}. )}
\label{fig:leakage}
\end{figure*}

We show the snapshots of the gas density and electromagnetic energy flux at $t=35,000 r_g/c$ for the a09rc0 and a09rc50 models side by side in Figure \ref{fig:leakage}. These snapshots demonstrate the characteristic behavior of magnetic fluxes during the quasi-equilibrium phase. 
It is apparent that model a09rc0 has a highly nonaxisymmetric accretion flow structure. Most strikingly, one can see from panels (b1) and (b2) that even along the midplane close to the BH gas does not flow inward along all directions as seen in typical disks, and there is a large azimuthal angle range persistently open for the gas to flow all the way out. The electromagnetic energy fluxes in these outflowing regions are large, indicating that such regions provide a channel for magnetic flux to leak out from the vicinity of the BH. We suspect that these ``magnetic bubbles'' are carried out by the wind or leak out due to buoyancy, which leads to a low level of magnetic flux present near the black hole. Model a09rc10 exhibits similar behavior to model a09rc0 (not shown). In contrast, model a09rc50 produces a steadier and more axisymmetric disk structure. There are still outflowing regions at high latitudes. However, along the midplane the outflows are patchy and not strong enough to disrupt the inflow. It is very probable that some magnetic bubbles still form and flow out for a short range, but later mix with the inflow and get shredded due to gas rotation. The magnetic flux is then brought back by the inflow and supplied to the vicinity of the BH, which allows the magnetic flux to accumulate and sustain the MAD state.
More careful analysis of the interaction of the magnetic flux with gas, which is out of the scope of this paper, will be needed to fully understand the physics here.

We also make an analysis to understand the roles played by the magnetic field and its coherent and turbulent components in transporting the gas angular momentum. While the details are given in Appendix \ref{sec:AM}, we give some highlights here. As in previous MAD simulations, we find that the Maxwell stress primarily produces viscosity in such disks. The turbulent component of the magnetic field dominates toward the disk midplane, suggesting that magnetohydrodynamic instabilities could be at work to transport angular momentum. However, along the disk surface and in the wind region, it is found that the coherent component of the magnetic field is more effective in transporting angular momentum. This effect is more dramatic in the a09rc50 run, which forms a more nonisotropic structure and produces stronger winds.

\section{Discussion}
\label{sec:discussion}

\subsection{Comparison with Previous GRMHD Simulations of Bondi-like Accretion Flows}

A few GRMHD simulations of Bondi-like accretion flow have been performed recently with focuses different from ours. For example, \citet{Ressler.2021} conducted various simulations of Bondi accretion to study how the orientation between the magnetic field and BH spin axis affects the behavior of the accretion flow and the jet. One of their simulations with the magnetic field aligned with the BH spin axis resembles our a09rc0 model. For this simulation, they reported a jet power higher than that in our a09rc0 model in the quasi-steady phase. 
We note that their simulation and ours have different initial gas and magnetic field configurations. More importantly, the accretion flows in the two simulations evolve for significantly different periods, likely leading to the discrepancy between the results. The \citet{Ressler.2021} simulation ran up to only $t \sim 20,000\ r_g/c$, during which period the accretion flow in our a09rc0 model is also in the MAD state, while the magnetic flux decreases to below MAD level in later phases and the jet power declines with that. More analysis is shown in Appendix \ref{sec:phiH}.

Very recently, as this letter was being prepared, \citet{Lalakos.2022} conducted a similar simulation of magnetized Bondi-like accretion flow with low specific angular momentum around a rapidly spinning BH. The gas flow in their simulation has a maximum specific angular momentum corresponding to a circularization radius $R_c=30 r_g$, which is between those of our a09rc10 and a09rc50 models. Their model also has a different initial magnetic field configuration by threading the gas with relatively weak magnetic fields outside the Bondi radius. Despite the differences, the accretion flow in their simulation evolves in a similar way to that in our a09rc50 model and also enters and sustains the MAD state. The combination of their results and ours suggests that the critical specific angular momentum for the accretion flow to achieve the MAD state corresponds to a circularization radius between $10r_g$ and $30 r_g$.

Furthermore, \citet{Nicholas.2022} studied BHL magnetized accretion onto a BH traveling through a uniform magnetized medium. Their B100R53 model has the most analogous configuration to our a09rc0 model with similar BH spin parameters. In addition, this run has a magnetic field strength in the wind similar to that in our initial disk. Their simulation produces a sporadic jet, which is in good accordance with our results. 
While their simulations with a stronger initial magnetic field in the wind can lead to sustained MAD accretion flow, there are some fundamental differences between their settings and ours: First, there is a steady supply of magnetic flux from the medium that the BH is traveling through in their simulations. Second, the relative motion between the BH (and its accretion disk) and the surrounding medium might help retain the magnetic flux in a similar way to that in our a09rc50 model.

\subsection{Applications to Various BH Accretion Systems}

In this section, we will briefly discuss how our results are relevant for a few BH accretion systems in which the accretion flows are believed to have very low specific angular momenta.

\textbf{\textit{Sgr A*}}. 
Sgr A*, the supermassive BH (SMBH) in the nucleus of our Galaxy, has a very low luminosity  $L_{\rm bol}\lesssim \ 10^{36} \text{erg}\ \text{s}^{-1}$ \citep{Narayan.1998, Bower.2019.ALMA}, indicating that it hosts a radiatively inefficient accretion flow \citep[see a review by][]{Yuan.2014}.  Linear polarization measurements of Sgr A* also constrain the accretion rate to as low as $10^{-9}-10^{-7} M_{\odot} \text{yr}^{-1}$ near the horizon \citep[e.g.][]{Marrone.2006}.
The accreting gas is likely supplied by the stellar wind from the neighbouring Wolf--Rayet stars, which gives a low cumulative angular momentum \citep[e.g.,][]{Quataert.2004}. Continuous accretion of stellar winds together with their magnetic fields is sufficient to form a MAD flow as shown by  numerical simulations \citep{Ressler.2020}. Furthermore, the recent Event Horizon Telescope observation of Sgr A* favors the model that Sgr A* is rapidly spinning and surrounded by a prograde accretion disk \citep{EHT.2022}. 

The findings above imply that our simulations of magnetized, Bondi-like accretion flow are very relevant for interpreting the properties of Sgr A*. For example, suppose we take the accretion rate of $10^{-9}$--$10^{-7} M_{\odot}\ \text{yr}^{-1}$ constrained from observation and use the minimum jet efficiency of about $10\%$ from our simulated accretion flow with zero angular momentum; we obtain a total jet power of $10^{36}$--$10^{38} \ \rm {erg \ s^{-1}}$, high enough to power Sgr A* radio and X-ray emission, which has an isotropic luminosity of at most $\sim10^{35}\ \text{erg}\ \text{s}^{-1}$ \citep{Morris.1996,Baganoff.2003}. Future higher-resolution observations and simulations/modeling of this type can allow us to better constrain the accretion flow structure and jet properties in Sgr A*.

\textbf{\textit{Long GRBs}}.
The emergence of long GRBs is usually explained using the collapsar model. In a collapsar, the core of a massive star has collapsed and formed a spinning BH, while the stellar envelope material is still accreting onto the BH, which launches a powerful, relativistic jet \citep{Woosley.1993}. However, there is debate whether the envelope of the progenitor star could have insufficient angular momentum to form a normal accretion disk \citep[e.g.][]{MacFadyen.1999}. Interestingly, our simulations show that even nonrotating accretion flows can launch a jet that is powerful enough to be the central engine of a long GRB, which alleviates the problem, as long as the BH is able to gain a large spin during the initial collapsing process.

\textbf{\textit{TDEs}}. 
In TDEs, a star is tidally disrupted by an SMBH along a parabolic orbit. Interestingly, the circularization radius of the stellar debris is usually a few tens of $r_g$ for typical stellar and BH parameters \citep{Rees.1988}. Therefore, our work shows that TDEs' accretion flows have the critical angular momentum needed to form MADs, if sufficient magnetic fields are supplied to the gas. Indeed, three TDEs have been observed to produce highly beamed X-rays, which are believed to be powered by jets \citep[e.g.,][]{Bloom11, Zauderer13, Tchekhovskoy14}, which is consistent with our prediction that accretion flow with such angular momentum can still reach the MAD condition and launch strong, relativistic jets. The majority of TDEs, however, produce thermal-like emissions likely associated with the disks or winds \citep{Dai18, Gezari21review}. This suggests either that most dormant SMBHs do not possess large spins or that most TDEs lack sufficient magnetic fluxes to form MADs.

\textbf{\textit{Wind-fed BH HMXBs}}.
HMXBs are often wind-fed systems instead of systems going through Roche-lobe overflow. In such a system, the accretion flow around the compact object forms from the capture of the wind from the donor giant star and, therefore, likely has a low specific angular momentum  \citep{Shapiro.1976}. Interestingly, all dynamically confirmed BH-HMXBs have BH spin values near the maximum \citep[e.g.][]{Liu.2008,Gou.2009}. In addition, collimated, relativistic jets have been observed from some BH-HMXBs \citep[e.g., Cyg X--1, ][]{Stirling.2001}. While the full hydrodynamics of wind-fed accretion can be complicated, our simulations of simplified Bondi-like accretion show that indeed powerful jets can be formed from these systems if they have sufficient magnetic fluxes. Future observations of the jet power might even allow us to constrain the spins of BHs in HMXBs.

\section{Summary}
In this work, we have performed novel 3D GRMHD simulations of a magnetized Bondi-like accretion flow around a rapidly spinning BH. The three models have different initial specific angular momenta ranging from zero to a small value corresponding to a Keplerian disk circularized at $50 r_g$. Our results highlight the role that initial gas angular momentum plays in forming MADs and producing powerful jets. We summarize the main findings as follows:

\begin{enumerate}
    \item For a Bondi-like magnetized accretion flow around a spinning BH, the gas specific angular momentum needs to reach a critical value so that a steady inflow structure can be formed, the MAD state can be reached and sustained, and a very powerful jet can be launched. This threshold of specific angular momentum is very low, which corresponds to a circularization radius likely between $10 r_g$ and $30 r_g$. 

    \item When the gas specific angular momentum does not reach this critical value, the accretion inflow is unsteady even close to the BH. In fact, the behavior of such flow is very different from the standard concept of accretion disks. If the BH spins fast, it drags the nearby gas to rotate with it. As a result of the conservation of angular momentum, the gas flow has different rotation directions at different inclinations. Furthermore, even around the midplane, the gas can flow in along certain azimuthal directions but persistently flow out to large radii at other azimuthal directions. We find that these outflow regions provide a channel for a large amount of magnetic flux to leak out. As a result, such accretion flow cannot sustain the MAD state. 
    
   \item It is further worth noting that even when the ``accretion flow'' is nonstandard as described above, it can still magnetically launch a relativistic jet via the Blandford--Znajek process. For example, we find that even a Bondi accretion flow with zero gas angular momentum around a rapidly spinning BH can launch a relativistic jet. However, the jet tends to be intermittent, wobbling, and relatively weak. The jet power can reach about 10\% of the accretion power. 
   
   \item In contrast, when the threshold of specific angular momentum has been reached, a steady gas inflow structure can form around the BH with coherent rotation and persistent inflow along the midplane, which resembles the accretion disks seen in standard GRMHD simulations using torus structures as the initial conditions. Under such conditions, we find that fewer magnetic fluxes leak out, which is likely because the rotating gas can shred outflowing magnetic bubbles and bring the magnetic flux back to the vicinity of the BH. This allows the MAD condition to be reached and robustly sustained.
   
   \item Under the circumstances that the gas has specific angular momentum above the low threshold and the BH spins fast, the jet efficiency from such magnetized accretion flow can approach and sometimes even exceed 100\%. 
  
\end{enumerate}

We emphasize that we have only tested the simplified scenario where the gas angular momentum, BH spin and magnetic field axes are all aligned, while they might be misaligned in some astrophysical systems. Furthermore, the treatment of cooling and radiative processes can be improved by conducting simulations using codes incorporating radiative transfer physics, which is particularly important in the super-Eddington accretion regime. 
Nonetheless, this work has disclosed that accretion flows around spinning BHs only need to possess small specific angular momenta to become MADs and launch very powerful jets. It also illustrates that the initial gas angular momentum and configuration can affect the results of GRMHD simulations of BH accretion.
Our results have important consequences for understanding various BH accretion systems with low gas angular momentum, such as Sgr A* and other low-luminosity AGNs, TDEs, GRBs and certain HMXBs, and, in particular, what sets off the jet power in such systems.\\

We thank S. Woosley for useful discussions at the early stages of the project. We also thank P. Sukova, A. Janiuk, D. Proga, L.L. Thomsen and the referee for helpful suggestions and comments.
T.M.K. and L.D. acknowledge support from the National Natural Science Foundation of China (HKU12122309) and the Hong Kong Research Grants Council (HKU27305119 and 17314822). AT acknowledges support by NSF grants AST-2009884, AST-2107839, AST-1815304, AST-1911080, OAC-2031997, and AST-2206471. We acknowledge support by the NSF through resources provided by NICS Kraken, where simulations were carried out, NICS Nautilus and TACC Frontera \citep{stanzione2020frontera}, where data were analyzed, and NCSA MSS and TACC Ranch, where data were backed up, under grants TG-AST100040 (TeraGrid), AST20011 (LRAC), and AST22011 (Pathways).
We acknowledge computational support from the high-performance computing facilities offered by the Information Technology Services Department at the University of Hong Kong and the Tianhe-2 supercluster.

%





\appendix
\counterwithin{figure}{section}

\twocolumngrid
\section{Simulation initial setup and parameters}\label{sec:initial}

\begin{figure*}[t!]
    \hspace*{-2.0cm}
    \begin{tabular}[b]{@{}p{0.3\textwidth}@{}}
        \gridline{\fig{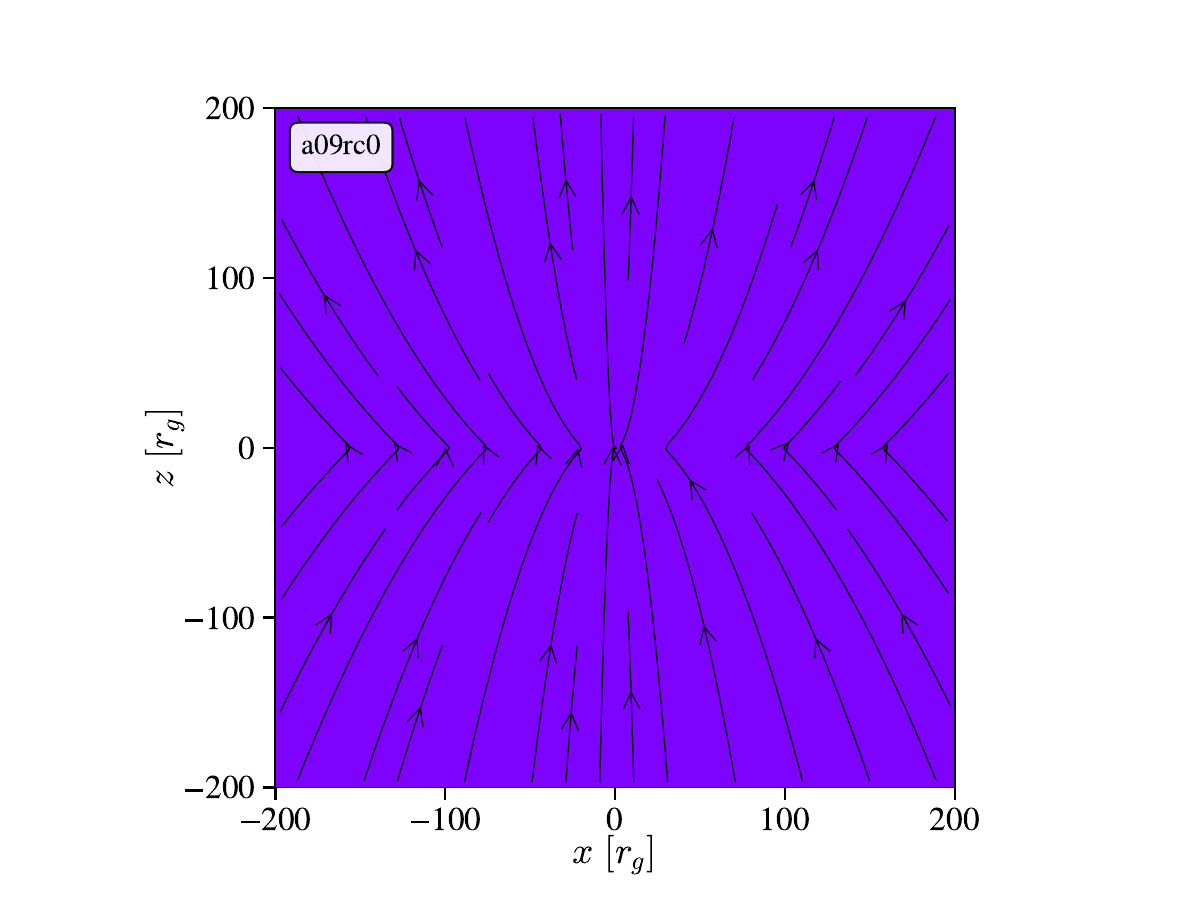}{0.47\textwidth}{} \label{fig:initialL_rc0}}
    \end{tabular}
    \hspace*{-1.2cm}
    \begin{tabular}[b]{@{}p{0.3\textwidth}@{}}
        \gridline{\fig{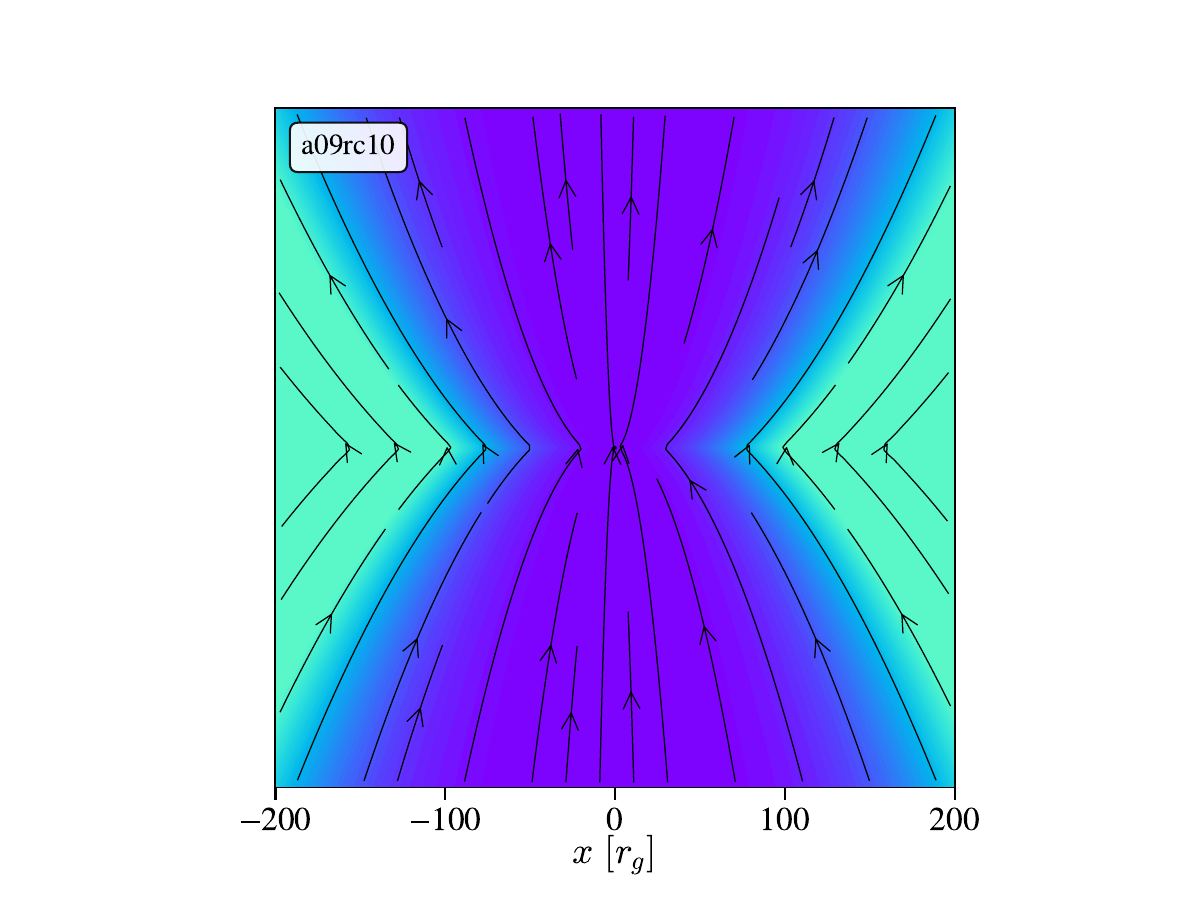}{0.47\textwidth}{} \label{fig:initialL_rc10}}
    \end{tabular}
    \hspace*{-0.8cm}
    \begin{tabular}[b]{@{}p{0.3\textwidth}@{}}
        \gridline{\fig{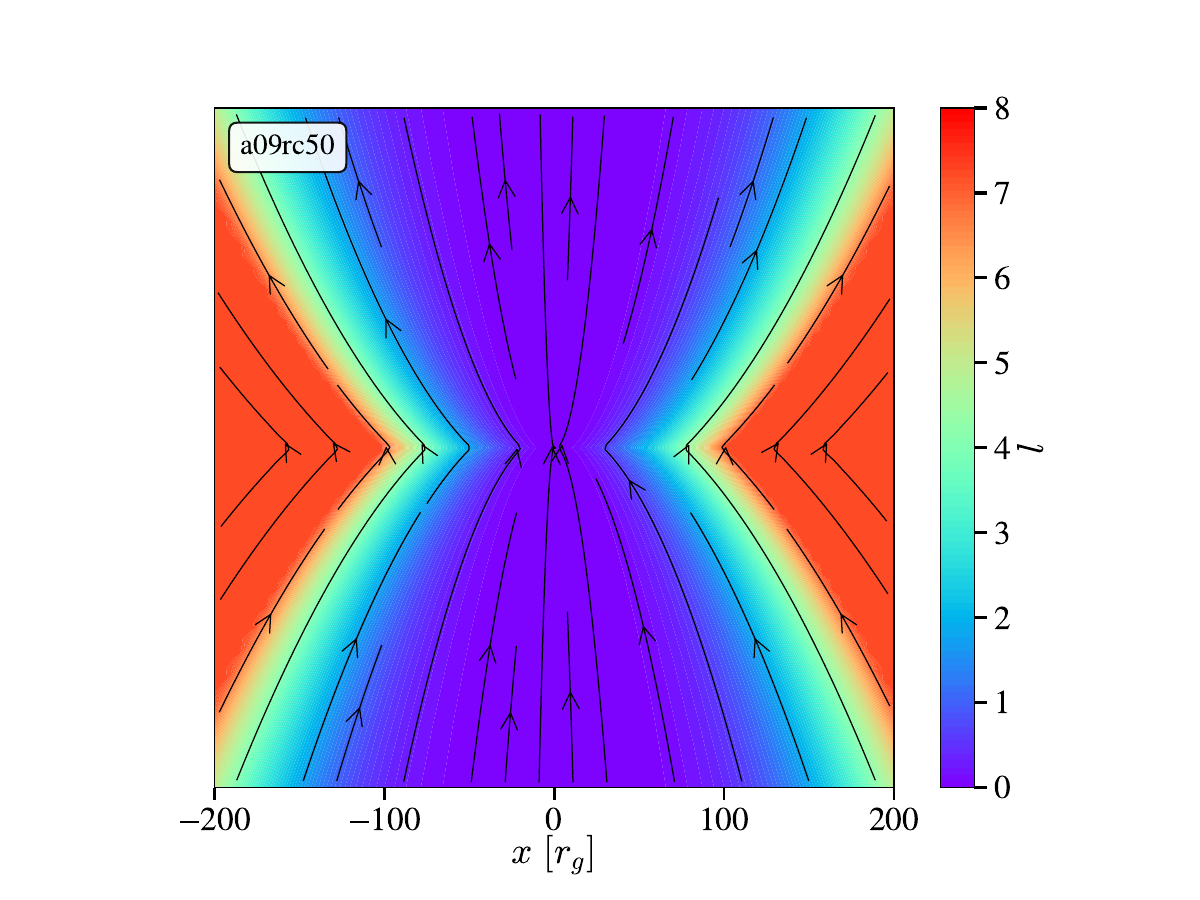}{0.47\textwidth}{} \label{fig:initialL_rc50}}
    \end{tabular}
    \vspace*{-1.0cm}
\caption[]{The initial conditions of the gas specific angular momentum $l$ and magnetic field of the three runs: (a) a09rc0, (b) a09rc10, (c) a09rc50. The colors show the initial specific angular momentum. The black lines indicate the initial magnetic field lines.}
\label{fig:initialL}
\end{figure*} 

We conduct three simulations with different initial gas specific angular momenta. In each of the simulations, a09rc0, a09rc10, and a09rc50 (Table~\ref{tab:summary}), the maximum specific angular momentum of the gas is set by the  circularization radius (Equation (\ref{eq: angular_momentum})): $R_c = 0$, $10 r_g$, and $50 r_g$, respectively. For our chosen magnetic flux distribution, a magnetic field line passing through a point $(r,\theta)$, with the magnetic flux function given by $\Psi=\Psi \left(r,\theta\right)$, will cross the equatorial plane, $\theta = \pi/2$, at the ``footpoint'' radius,
\begin{equation}
R_{\rm fp}(r,\theta)  = r (1-|\cos\theta|).
\label{eq:footpoint}
\end{equation}
The initial specific angular momentum distribution is set as 
\begin{equation}
    l=
    \begin{cases}
    l_{\rm max}                                       & \text{for} \ R_{\rm{}fp}>R_{\rm{}solid}, \\
    l_{\rm max} \times R_{\rm fp}^2 /R_{\rm{}solid}^2  & \text{for}\ R_{\rm{}fp}\leq R_{\rm{}solid}, 
    \end{cases} \label{eq: angular_momentum_full}
\end{equation} 
where $R_{\rm{}solid}\equiv100 r_g$ and $R_{\rm fp}$ is given by Equation (\ref{eq:footpoint}). Using this setup, close to the BH, gas rotates approximately as a solid body (i.e., $u^{\phi}$ is roughly constant), so we avoid having a nonphysically high specific angular momentum. Also, outside the equator, the value of $l$ is propagated along the magnetic field lines. We plot the initial specific angular momentum, together with the magnetic field lines, in Figure \ref{fig:initialL}.

A few numerical and physical parameters of the simulations are listed in table \ref{tab:summary}. Besides the various parameters discussed in the main text, we show the value of $s$, the BH dimensionless spin-up parameter, which is defined as
\begin{equation}
    s \equiv \frac{{\rm d} j}{{\rm d} t} \frac{M_{\rm{}BH}}{\langle\dot{M}_H \rangle_t} = -j - 2a(1-\eta)
\end{equation}
where $j=\frac{\int T^r_{\phi} \rm dA_{\rm \theta\phi}}{\langle\dot{M}_H \rangle_t}$ is the specific angular momentum flux of the BH. Positive values indicate the spin-up of the BH. We can see in all three runs, the jet has extracted rotation energy from the BH.

\section{Conversion from geometric units to CGS units}\label{sec:convert}

We express all the important quantities reported in this letter in Table \ref{tab:unit-conversion} and show the conversion from geometric units to CGS units. As nonradiative GRMHD simulations are scale-invariant, two basic scaling parameters are needed: (1) the BH mass $M_{\rm BH}$, for which we use a scaling BH mass $M_0=10^6 M_\odot$, and (2) the density $\tilde{\rho}$, for which we use a scaling density $\rho_0 = 10^{-10}\ \rm g\ cm^{-3}$. 

\begin{deluxetable*}{cccc} 
\label{tab:unit-conversion}
\tablecaption{Summary of Unit Conversion of Nonradiative Ideal GRMHD Quantities}
\tablewidth{0pt}
\tablehead{
\colhead{\begin{tabular}{@{}c@{}} Scaling \\ Parameters\end{tabular}} &   \colhead{\begin{tabular}{@{}c@{}}Geometric Units of the \\ Corresponding Parameters\end{tabular}} &  \colhead{Conversion to CGS Units}   &  \colhead{Definition} 
}
\startdata
$\mathcal{L}$ ($r_g$)	&   $GM_{\rm BH}/c^2$             &  $\times 1.48\times10^{11} \times (M_{\rm BH}/ M_0)$ cm   &   Length Scale   \\
$\mathcal{T}$	&   $\mathcal{L}/c$             &  $\times 4.93 \times (M_{\rm BH}$ /$M_0$) s   &  Time Scale 	 \\
$\mathcal{V}$	&   $c$                           &    $\times 3.00\times10^{10}\ \rm cm\ s^{-1}$                  &   Velocity Scale 	 \\
$\tilde{\rho}$  &   Arbitrary                       &    $\times \rm (a\ chosen\ number)\ g\ cm^{-3}$                                &   Density Scale \\
$\mathcal{M}$   &   $\tilde{\rho} \mathcal{L}^3$    &   $\times 3.21\times10^{23} \times (\tilde{\rho}/\rho_0) \times (M_{\rm BH}/M_0)^3$ g  &   Mass Scale \\
$\dot{\mathcal{M}}_H$     &   $\tilde{\rho} \mathcal{L}^2 c$    &   $\times 6.53\times10^{22} \times (\tilde{\rho}/\rho_0) \times (M_{\rm BH}/M_0)^2 \ \rm g\ s^{-1}$    &   Mass Accretion Rate Scale\\
$\mathfrak{l}$             &   $GM_{\rm BH}/c$             &   $\times 4.43\times10^{21} \times (M_{\rm BH}/M_0)\ \rm cm^2\ s^{-1}$    &   Specific Angular Momentum Scale  \\
$\mathcal{U}$   &   $\tilde{\rho} \mathcal{V}^2$    &   $\times 9 \times 10^{30} \times (\tilde{\rho}/\rho_0) \ \rm erg \ cm^{-3}$     &   Energy Density Scale \\
$\mathfrak{b}$     &   $c\sqrt{2\tilde{\rho}}$                 &   $\times 4.24 \times10^{5} \times \sqrt{\tilde{\rho}/\rho_0} \ \rm G\ cm^2$              &   Magnetic Flux Scale \\
\enddata
\end{deluxetable*}

\section{Resolving and Suppressing MRI} \label{sec:mri}

\begin{figure}
\centering
\includegraphics[width=\columnwidth]{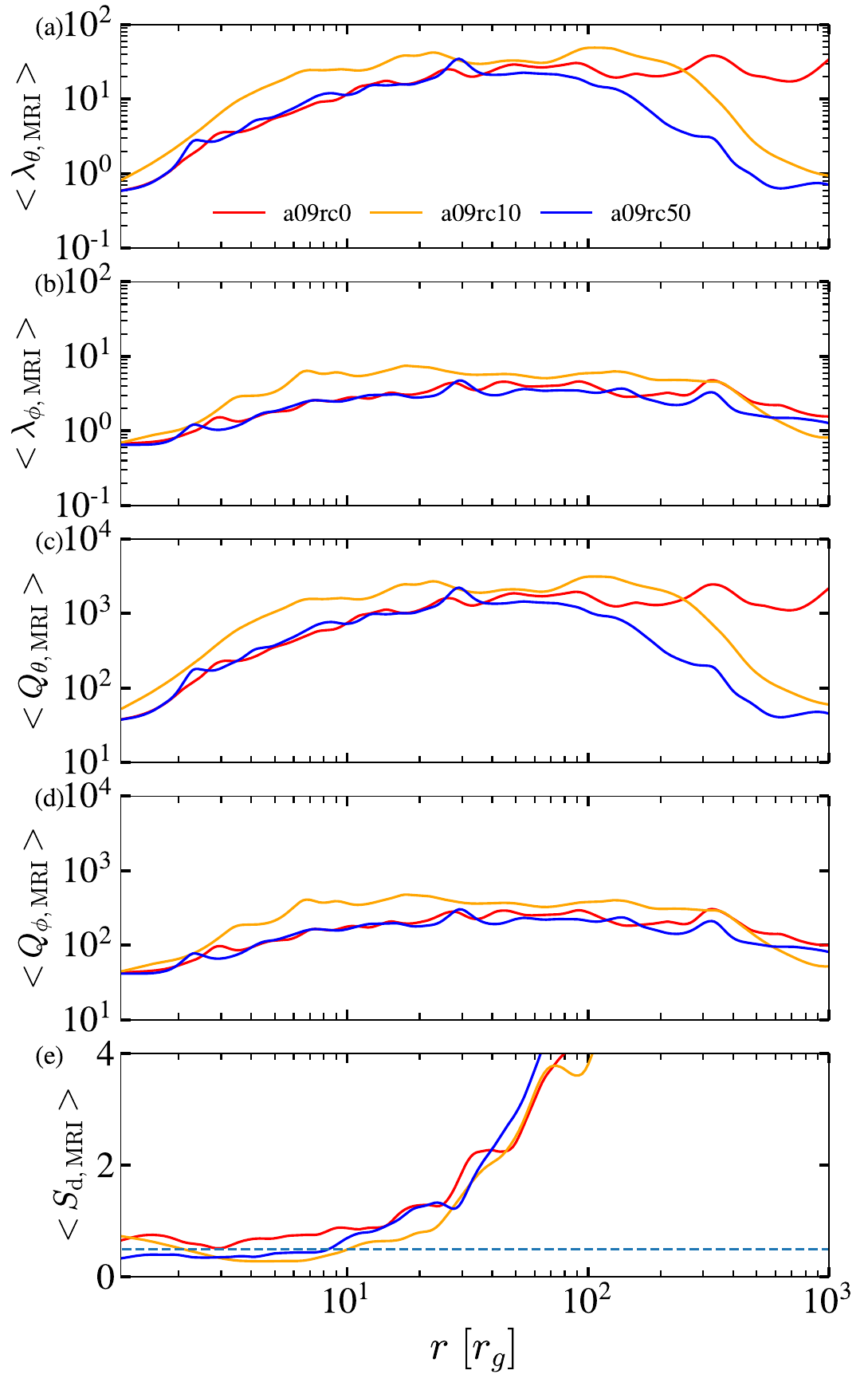}
\caption{The time-$\theta$-$\phi$-averaged radial profiles of (a) the fastest-growing MRI wavelength in the $\theta$-direction, $\lambda_{\rm \theta, MRI}$, (b) the fastest-growing MRI wavelength in the $\phi$-direction, $\lambda_{\rm \phi, MRI}$, (c) the quality factor in the $\theta$-direction, $Q_{\rm \theta, MRI}$, (d) the quality factor in $\phi$-direction, $Q_{\rm \phi, MRI}$, and (e) the MRI suppression factor, $S_{\rm disk, MRI}$. The colors indicate the three models as in Figure \ref{fig:radial_profile}. The dashed line indicates the criterion that the MRI is suppressed in the disks if $S_{\rm disk, MRI} \lesssim 1/2$.}
\label{fig:MRI}
\end{figure}

MRI is an important process for angular momentum transfer in accretion flow. Its fastest-growing wavelength is
\begin{equation} \label{MRI}
    \lambda_{x, \rm MRI} \approx 2\pi \frac{|v_{x,\rm A}|}{|\Omega|}
\end{equation}
with $x=\theta$ or $\phi$ \citep{Hawley.1995}. Here, $|v_{x, \rm A}| = \sqrt{b_x b^x/(b^2+\rho+u_g+p_g)}$ is the Alfv\'en speed in the \textit{x}-direction. 
We plot the time-$\theta$-$\phi$-averaged $\lambda_{\theta, \rm MRI}$ and $\lambda_{\phi, \rm MRI}$ in Figures \ref{fig:MRI}(a) and (b), respectively. 
Note that we smooth all the quantities in Figure \ref{fig:MRI} using a second-order Whittaker–-Henderson smoothing method with a smooth factor of $100$ to make the plots more readable. 

Furthermore, we check the number of $\theta$-grid and $\phi$-grid cells per fastest-growing MRI mode, namely the MRI quality factor, 
\begin{equation} \label{Q_MRI}
    Q_{x, \rm MRI} = \frac{\lambda_{x, \rm MRI}}{\Delta_{x}}
\end{equation}
with $x=\theta$ or $\phi$, where $\Delta_{x}$ is the $x$-grid cell length. If $Q_{x, \rm MRI}>6$, the MRI is considered resolved. 
We plot $Q_{\rm \theta, MRI}$ and $Q_{\rm \phi, MRI}$ in Figure \ref{fig:MRI}(c) and (d), respectively. It is apparent that the MRI is well resolved in the $\theta$- and $\phi$- direction in the whole simulation domain in all three models. 

Lastly, we check if the MRI is suppressed in the disk region. The MRI suppression factor can be calculated as the number of MRI wavelengths across the smallest dimension of the disk:
\begin{equation} \label{S_MRI}
    S_{\rm disk, MRI} = \frac{2r(H/R)}{\lambda_{\rm \theta, MRI}} .
\end{equation}
The disk is magnetorotationally stable if half of the fastest-growing MRI wavelength cannot fit within the full disk, which sets the criterion that the MRI is suppressed in the disks if $S_{\rm disk, MRI}<1/2$ \citep{Balbus.1998, McKinney.2012}. We plot $S_{\rm disk, MRI}$ in Figure \ref{fig:MRI}(e). One can see that in models a09rc10 and a09rc50 the MRI is mildly suppressed in the inflow region within $r\lesssim 10r_g$. However, in model a09rc0, the MRI process can survive, although only barely, in the inflow near the BH. 

\section{Magnetic fields in the inflow and outflow} \label{sec:Magnetic_field}
\begin{figure}[ht]
\centering
\includegraphics[width=1.05\columnwidth]{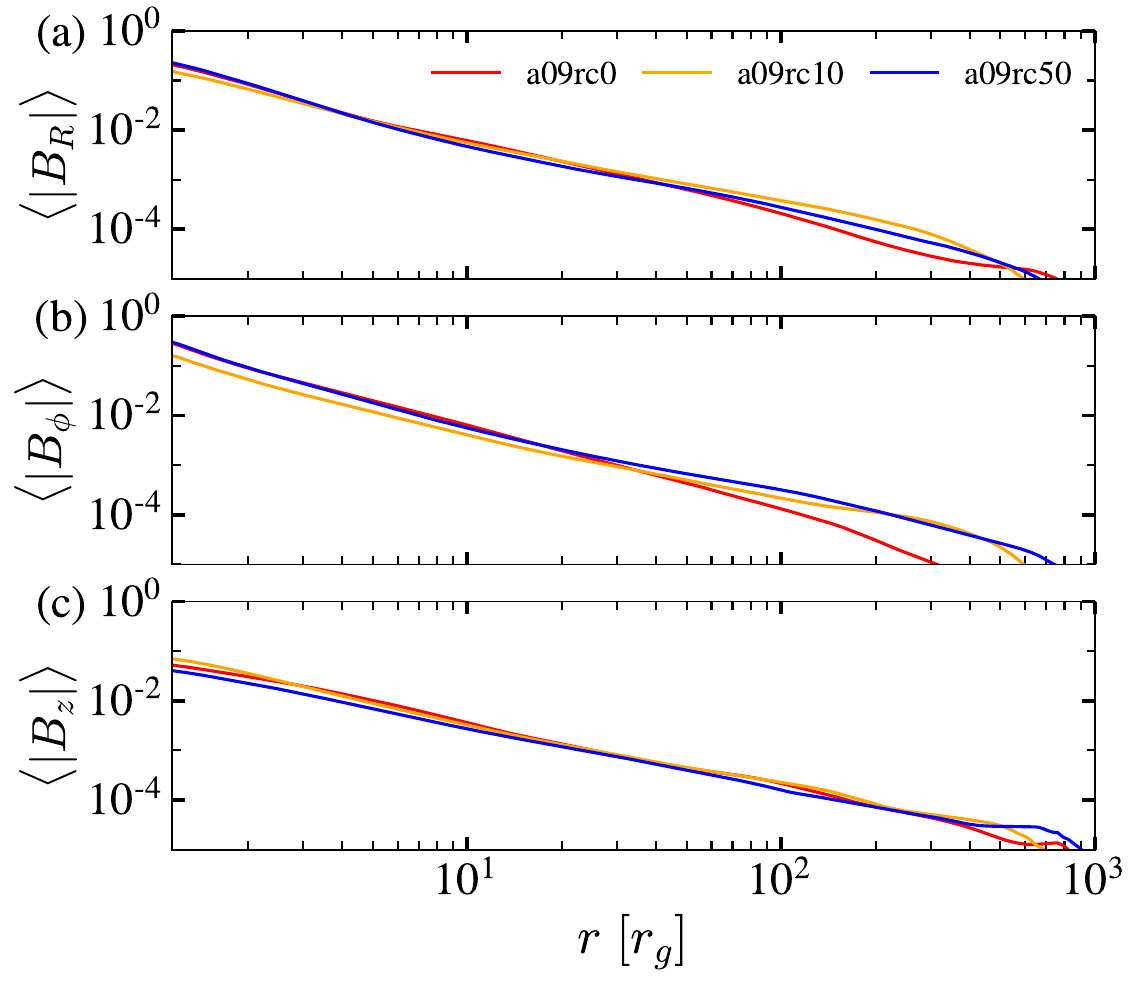}
\caption{The time and angle-integrated radial profiles of (a) $|B_R|$, (b) $|B_{\phi}|$, and (c) $|B_z|$ of the inflow. The colors indicate the three models as in Figure \ref{fig:radial_profile}. We find the three magnetic field components evolve similarly in the inflow equilibrium regions ($r_g\lesssim 100 r_g$) in all three models. }
\label{fig:B-field}
\end{figure}

One can calculate the `quasi-orthonormal' lab-frame magnetic field three-vectors as
\begin{equation} \label{Bfield-orthonormal}
    \begin{split}
        B_r &=  B^r \sqrt{g_{rr}}, \\
        B_{\theta}  &=  B^{\theta} \sqrt{g_{\theta\theta}}, \\
        B_{\phi} &= B^{\phi} \sqrt{g_{\phi\phi}}. \\
    \end{split}
\end{equation} 
One can further express these vectors in cylindrical coordinates,
\begin{equation} \label{Bfield-cylindrical}
    \begin{split}
        B_R &=  B_r \sin\theta + B_{\theta}\cos\theta , \\
        B_z &= B_r \cos\theta - B_{\theta}\sin\theta . \\
    \end{split}
\end{equation} 

Figure \ref{fig:B-field} shows the time and angle-integrated radial profiles of the magnetic field three-vectors: (a) $|B_R|$, (b) $|B_{\phi}|$, and (c) $|B_z|$ of the inflow. Note that $B_R$ and $B_{\phi}$ have different signs above and below the midplane, so we focus on the absolute magnitude of the magnetic field strength.
In the three models, the magnetic fields evolve similarly in the inflow equilibrium region with $r_g\lesssim 100 r_g$. In this region, $|B_R|$ and $|B_{\phi}|$ decay approximately following $B \propto r^{-1.5}$, while $|B_z|$ declines more gently following $B \propto r^{-1.2}$.


\begin{figure}[ht]
\centering
\includegraphics[width=1.18\columnwidth]{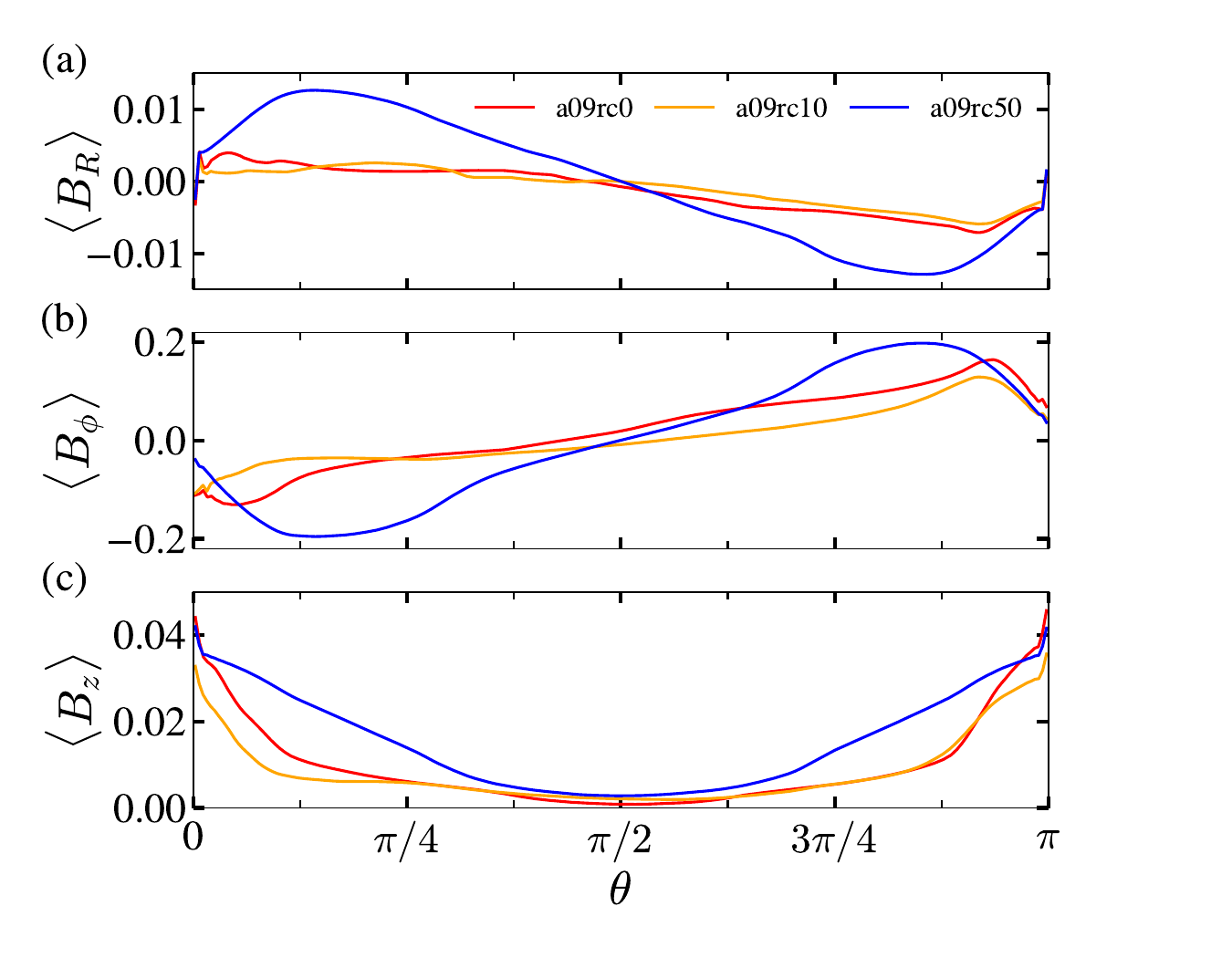}
\caption{The time and $\phi$-averaged $\theta$-profiles of (a) $B_R$, (b) $B_{\phi}$, and (c) $B_z$ along the surface at $r=6 r_g$. The colors indicate the three models as in Figure \ref{fig:radial_profile}. $B_R$ and $B_{\phi}$ have different signs above and below the midplane. }
\label{fig:B-field-angular}
\end{figure}

Figure \ref{fig:B-field-angular} shows the $\theta$-profiles of (a) $B_R$, (b) $B_{\phi}$, and (c) $B_z$, all evaluated on the surface at $r=6r_g$. These quantities are time/$\phi$-averaged. $B_R$ and $B_{\phi}$ have different signs above and below the midplane. We find that model a09rc50 generates the largest magnitude of $B_R$ and $B_{\phi}$ in the wind region.

\section{Angular momentum transport in the accretion flow} \label{sec:AM}
\begin{figure*}[ht]
\centering
\includegraphics[width=0.68\textwidth]{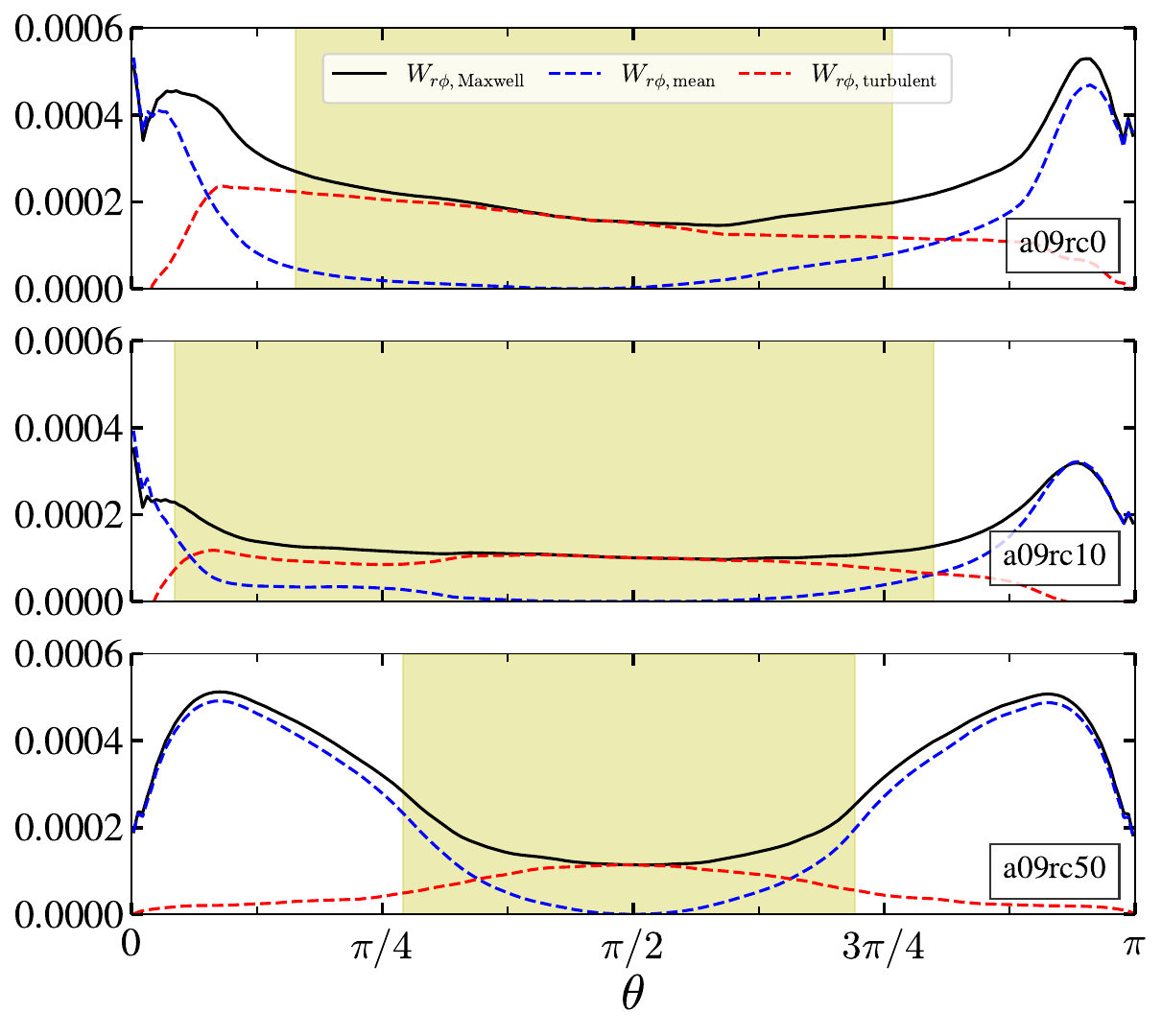}
\caption{The time and $\phi$-averaged $\theta$-profiles of the total Maxwell stress $W_{r\phi,\rm Maxwell}$ (solid lines), mean-field torque stress $W_{r\phi,\rm mean}$ (blue dashed lines), and turbulent field torque stress $W_{r\phi,\rm turbulent}$ (red dashed lines) for the surface at $r=6 r_g$. The light yellow shading highlights the inflow region where $u^r<0$ at $r=6 r_g$. The Maxwell stress is always positive at all latitudes. In all models, the turbulent component of the Maxwell stress dominates the inflow region toward the midplane, and the mean-field component dominates the outflow region. } 
\label{fig:maxwell_stress}
\end{figure*}

The viscosity parameter $\alpha$ is defined as the ratio of the $r\phi$-component of the stress--energy tensor $W_{r\phi}$ to the sum of the gas pressure $p_g$ and magnetic pressure $p_b$,
\begin{equation}\label{alpha}
    \alpha = \frac{W_{r\phi}}{p_g+p_b},
\end{equation}
for which we have included the magnetic pressure in the denominator to keep $\alpha<1$ in all distances. Note that this choice makes $\alpha$ depend on the magnetic pressure very close to the BH ($\lesssim 10 r_g$), where the magnetic pressure is comparable to the gas pressure. 
The Reynolds stress $W_{r\phi, \rm Reynolds}$ and Maxwell stress $W_{r\phi, \rm Maxwell}$ contribute to the total stress $W_{r\phi}$. The definitions of the Reynolds and Maxwell stresses are ambiguous. We follow the expressions given by \citet{Krolik.2005} and \citet{Beckwith.2008}. Note that we first convert the quantities from the Kerr--Schild coordinate to the Boyer--Lindquist coordinate following their formulations.
Also, we measure the quantities in the fluid frame, since it is believed that MHD instabilities are primarily caused by local instabilities in the fluid.

The Maxwell stress comes from the EM-term energy--stress tensor ${T^{\rm EM}}^{r}_{\phi}$: 
\begin{equation} \label{maxwell_stress}
    \begin{split}
        {T^{\rm EM}_{\rm BL}}^{r}_{\phi} &=  b^2 \sqrt{g_{rr}} u^{r} \sqrt{g^{\phi\phi}} u_{\phi} - \sqrt{g_{rr}} b^{r} \sqrt{g^{\phi\phi}} b_{\phi}. \\
    \end{split}
\end{equation} 
In particular, we are more interested in the second component $-\sqrt{g_{rr}} b^{r} \sqrt{g^{\phi\phi}} b_{\phi}$, which is the magnetic torque term responsible for transferring angular momentum outward from the gas. This stress component can be generated by field turbulence or large-scale magnetic fields, and it can be nonzero even in laminar flow. We can therefore decompose this term into a mean-field (coherent) component and a turbulent field component as
\begin{equation} \label{maxwell_torque_stress}
    \begin{split}
        W_{\rm r\phi, Maxwell} &=  W_{\rm r\phi, mean} + W_{\rm r\phi, turbulent} \\
        W_{\rm r\phi, mean} &= -\langle \sqrt{g_{rr}} b^{r} \rangle \langle \sqrt{g^{\phi\phi}} b_{\phi} \rangle \\
        W_{\rm r\phi, turbulent} &= - \langle \sqrt{g_{rr}} b^{r} \sqrt{g^{\phi\phi}} b_{\phi} \rangle + \langle \sqrt{g_{rr}} b^{r} \rangle \langle \sqrt{g^{\phi\phi}} b_{\phi} \rangle .
    \end{split}
\end{equation} 


The relevant Reynolds stress--energy tensor is calculated as
\begin{equation} \label{reynolds_stress}
    \begin{split}
        W_{\rm r\phi, Reynolds} &=  \rho \sqrt{g_{rr}} \delta u^r \sqrt{g^{\phi\phi}} \delta u_{\phi} \\
    \end{split}
\end{equation} 
where $\delta u^{r}$ and $\delta u_{\phi}$ are the difference between $u^r$ and the time-azimuthally averaged $u^r$, $\delta u^r = |u^r - \langle u^r \rangle_{\phi}|$, and the difference between $u_{\phi}$ and the time-azimuthally averaged $u_{\phi}$, $\delta u_{\phi} = |u_{\phi} - \langle u_{\phi} \rangle_{\phi}|$.

We show the $\theta$-profiles of the Maxwell stress and its components, all evaluated on the surface at $r=6 r_g$, in Figure \ref{fig:maxwell_stress}.
These quantities are time and $\phi$-averaged. 
We highlight the profiles with light yellow in the inflow region where $u^r<0$ at $r=6 r_g$. 
In the three models, the Maxwell stress $W_{r\phi}$ is always positive at all latitudes, implying that magnetic fields in the disk, wind, and jet always transport gas angular momentum away.
In particular, in the a09rc50 model, the winds have much higher $W_{r\phi}$ than the inflows, which shows that they play a more important role in transporting angular momentum. 

In the inflow, we find that the turbulent component $W_{r\phi,\rm turbulent}$ always dominates the Maxwell stress in the dense inflow region toward the midplane, which suggests that angular momentum is still mainly transported via the magnetic field turbulence produced by magnetohydrodynamic instabilities in such regions. 
In the wind region, the mean-field component $W_{r\phi,\rm mean}$ always dominates the Maxwell stress.
Interestingly, in model a09rc50, where the inflow has formed a MAD, the mean-field component $W_{r\phi,\rm mean}$ in the disk surface layer is significantly larger than the turbulent component. Therefore, in this model, the magnetic torque generated by the large-scale mean magnetic field can drive the accretion of gas on the surface of the MAD. This is consistent with the findings by \citet{Mishra.2020}, who performed MHD simulations of geometrically thin, strongly magnetized disks (without winds or jets) and showed that the ordered magnetic fields drive accretion along the surface of such disks.

\begin{figure*}[ht]
\centering
\includegraphics[width=0.68\textwidth]{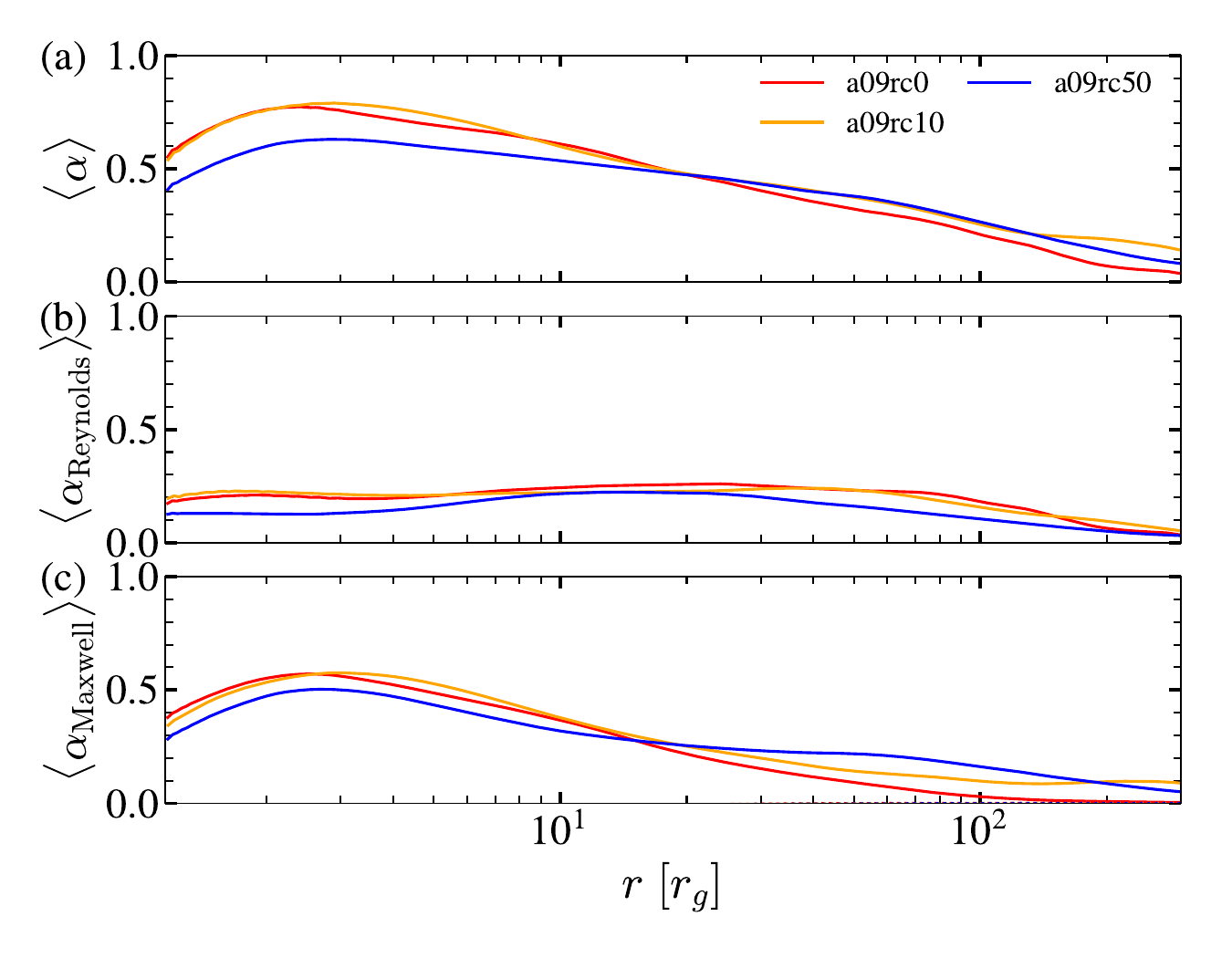}
\vspace*{-0.5cm}
\caption{The time and angle-integrated radial profiles of (a) the total disk viscosity parameter $\alpha$ of the accretion inflow and its components: (b) the Reynolds viscosity parameter $\alpha_{\rm Reynolds}$ and (c) the Maxwell viscosity parameter $\alpha_{\rm Maxwell}$. The colors indicate the three models as in Figure \ref{fig:radial_profile}. The Maxwell viscosity $\alpha_{\rm Maxwell}$ is relatively high in the inner inflow in all simulations, suggesting the Maxwell stress mainly causes the transport of angular momentum and accretion. } 
\label{fig:alpha_viscosity}
\end{figure*}

Next, in Figure \ref{fig:alpha_viscosity} we show the time/angle-integrated radial profiles of (a) the total disk $\alpha$ parameter of the accretion inflow and its components: (b) the Reynolds viscosity parameter $\alpha_{\rm Reynolds} = W_{r\phi,\rm Reynolds}/(p_g+p_b)$ and (c) the Maxwell viscosity parameter $\alpha_{\rm Maxwell} = W_{r\phi, \rm Maxwell}/(p_g+p_b)$. In the inner inflow at $r \lesssim 10 r_g$, the gas has a total $\alpha$ parameter of 0.4--0.8, and model a09rc50 has a smaller $\alpha$ than the other two models. The Reynolds viscosity parameter ($\alpha_{\rm Reynolds}$ $\sim$ 0.1--0.2) is smaller than the Maxwell viscosity parameter ($\alpha_{\rm Maxwell}$ $\sim$ 0.3--0.6) in all simulations. This means that the magnetic fields primarily drive the gas accretion in the inner disk. 
In the outer regions of the disk ($r \gtrsim 20 r_g$), the total disk $\alpha$ parameter gradually declines and becomes similar in all simulations, with $\alpha_{\rm Maxwell}$ decaying faster than $\alpha_{\rm Reynolds}$.

\section{Histogram of MAD parameter}
\label{sec:phiH}

\begin{figure*}[ht]
\gridline{\fig{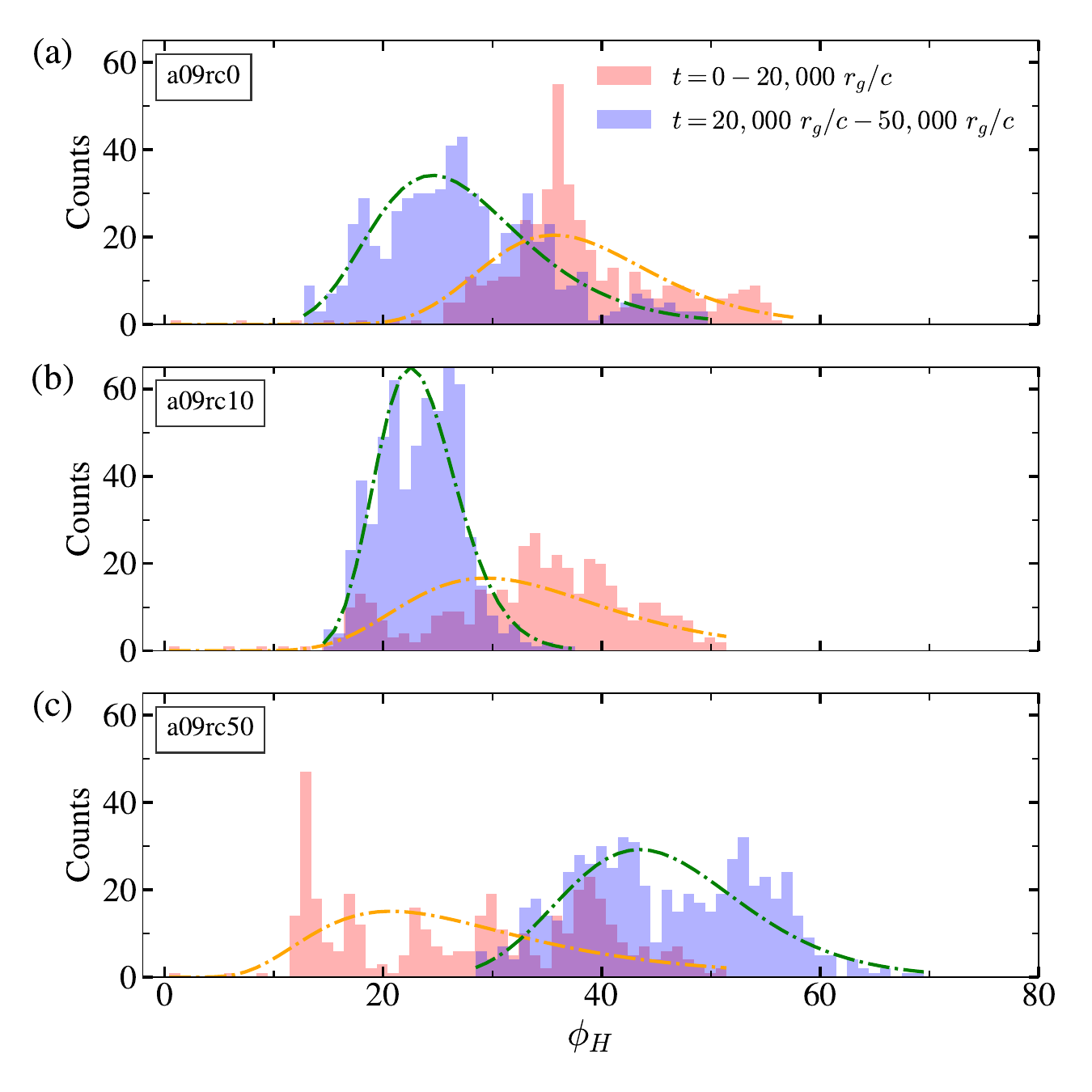}{0.7\textwidth}{}}
\vspace*{-1.2cm}
\caption{Histograms of the dimensionless magnetic flux around the horizon $\phi_H$ of the three models: (a) a09rc0, (b) a09rc10, and (c) a09rc50. We sample $\phi_H$ in two different time ranges: 1) the early phase $t=$ (0 -- 20,000)$r_g/c$, plotted in red, and 2) the late, quasi-equilibrium phase $t=$ (20,000 -- 50,000)$r_g/c$ (when we compute the average values of the quantities), plotted in blue. The orange and green dashed--dotted lines are the lognormal distribution functions fitted on the profiles of $\phi_H$ in the early and late phases, respectively. }
\label{fig:histogram}
\end{figure*}

Figure \ref{fig:histogram} shows histograms of the instant dimensionless magnetic flux $\phi_H$ in all snapshots of the three runs, either over the early phase of $t=$ (0--20,000)$r_g/c$ or over the late phase of $t=$ (20,000--50,000)$r_g/c$. The accretion flow has established inflow equilibrium out to $r \gtrsim 100 r_g$ only in the late phase, during which period we have evaluated the results. 

One can see that the accretion flow has distinct properties at the early and late phases. For models a09rc0 and a09rc10, the inner accretion flows are MAD or nearly MAD before $t\sim$ 20,000$r_g/c$ with $\phi_H$ varying around 40. However, the magnetic flux decays with time and the accretion flow is no longer MAD in the quasi-equilibrium phase. Model a09rc50 shows the opposite behavior. The MAD state is slowly developed and then maintained in the quasi-equilibrium phase.

These results demonstrate the importance of conducting GRMHD simulations for a sufficiently long duration and probably explain some differences between the results obtained by \citet{Ressler.2021} and this work.


\bibliography{export-bibtex}{}

\begin{thebibliography}{}
\expandafter\ifx\csname natexlab\endcsname\relax\def\natexlab#1{#1}\fi
\providecommand{\url}[1]{\href{#1}{#1}}
\providecommand{\dodoi}[1]{doi:~\href{http://doi.org/#1}{\nolinkurl{#1}}}
\providecommand{\doeprint}[1]{\href{http://ascl.net/#1}{\nolinkurl{http://ascl.net/#1}}}
\providecommand{\doarXiv}[1]{\href{https://arxiv.org/abs/#1}{\nolinkurl{https://arxiv.org/abs/#1}}}

\bibitem[{{Abramowicz} {et~al.}(1988){Abramowicz}, {Czerny}, {Lasota}, \&
  {Szuszkiewicz}}]{Abramowicz.1988}
{Abramowicz}, M.~A., {Czerny}, B., {Lasota}, J.~P., \& {Szuszkiewicz}, E. 1988,
  \apj, 332, 646, \dodoi{10.1086/166683}

\bibitem[{{Baganoff} {et~al.}(2003){Baganoff}, {Maeda}, {Morris}, {Bautz},
  {Brandt}, {Cui}, {Doty}, {Feigelson}, {Garmire}, {Pravdo}, {Ricker}, \&
  {Townsley}}]{Baganoff.2003}
{Baganoff}, F.~K., {Maeda}, Y., {Morris}, M., {et~al.} 2003, \apj, 591, 891,
  \dodoi{10.1086/375145}

\bibitem[{{Balbus} \& {Hawley}(1998)}]{Balbus.1998}
{Balbus}, S.~A., \& {Hawley}, J.~F. 1998, Reviews of Modern Physics, 70, 1,
  \dodoi{10.1103/RevModPhys.70.1}

\bibitem[{{Beckwith} {et~al.}(2008){Beckwith}, {Hawley}, \&
  {Krolik}}]{Beckwith.2008}
{Beckwith}, K., {Hawley}, J.~F., \& {Krolik}, J.~H. 2008, \mnras, 390, 21,
  \dodoi{10.1111/j.1365-2966.2008.13710.x}

\bibitem[{{Bisnovatyi-Kogan} \& {Ruzmaikin}(1974)}]{Bisnovatyi-Kogan.1974}
{Bisnovatyi-Kogan}, G.~S., \& {Ruzmaikin}, A.~A. 1974, \apss, 28, 45,
  \dodoi{10.1007/BF00642237}

\bibitem[{{Bisnovatyi-Kogan} \& {Ruzmaikin}(1976)}]{Bisnovatyi-Kogan.1976}
---. 1976, \apss, 42, 401, \dodoi{10.1007/BF01225967}

\bibitem[{{Blandford} \& {Znajek}(1977)}]{Blandford.1977}
{Blandford}, R.~D., \& {Znajek}, R.~L. 1977, \mnras, 179, 433,
  \dodoi{10.1093/mnras/179.3.433}

\bibitem[{{Bloom} {et~al.}(2011){Bloom}, {Giannios}, {Metzger}, {Cenko},
  {Perley}, {Butler}, {Tanvir}, {Levan}, {O'Brien}, {Strubbe}, {De Colle},
  {Ramirez-Ruiz}, {Lee}, {Nayakshin}, {Quataert}, {King}, {Cucchiara},
  {Guillochon}, {Bower}, {Fruchter}, {Morgan}, \& {van der Horst}}]{Bloom11}
{Bloom}, J.~S., {Giannios}, D., {Metzger}, B.~D., {et~al.} 2011, Science, 333,
  203, \dodoi{10.1126/science.1207150}

\bibitem[{{Bondi}(1952)}]{Bondi.1952}
{Bondi}, H. 1952, \mnras, 112, 195, \dodoi{10.1093/mnras/112.2.195}

\bibitem[{{Bower} {et~al.}(2019){Bower}, {Dexter}, {Asada}, {Brinkerink},
  {Falcke}, {Ho}, {Inoue}, {Markoff}, {Marrone}, {Matsushita}, {Moscibrodzka},
  {Nakamura}, {Peck}, \& {Rao}}]{Bower.2019.ALMA}
{Bower}, G.~C., {Dexter}, J., {Asada}, K., {et~al.} 2019, \apjl, 881, L2,
  \dodoi{10.3847/2041-8213/ab3397}

\bibitem[{{Dai} {et~al.}(2018){Dai}, {McKinney}, {Roth}, {Ramirez-Ruiz}, \&
  {Miller}}]{Dai18}
{Dai}, L., {McKinney}, J.~C., {Roth}, N., {Ramirez-Ruiz}, E., \& {Miller},
  M.~C. 2018, \apj, 859, L20, \dodoi{10.3847/2041-8213/aab429}

\bibitem[{{Davis} \& {Tchekhovskoy}(2020)}]{Davis.2020}
{Davis}, S.~W., \& {Tchekhovskoy}, A. 2020, \araa, 58, 407,
  \dodoi{10.1146/annurev-astro-081817-051905}

\bibitem[{{De Villiers} \& {Hawley}(2003)}]{DeVilliers.2003}
{De Villiers}, J.-P., \& {Hawley}, J.~F. 2003, \apj, 589, 458,
  \dodoi{10.1086/373949}

\bibitem[{Fryer(1999)}]{Fryer.1999}
Fryer, C.~L. 1999, The Astrophysical Journal, 522, 413, \dodoi{10.1086/307647}

\bibitem[{{Gammie} {et~al.}(2003){Gammie}, {McKinney}, \&
  {T{\'o}th}}]{Gammie.2003}
{Gammie}, C.~F., {McKinney}, J.~C., \& {T{\'o}th}, G. 2003, \apj, 589, 444,
  \dodoi{10.1086/374594}

\bibitem[{{Gezari}(2021)}]{Gezari21review}
{Gezari}, S. 2021, \araa, 59, \dodoi{10.1146/annurev-astro-111720-030029}

\bibitem[{{Gou} {et~al.}(2009){Gou}, {McClintock}, {Liu}, {Narayan}, {Steiner},
  {Remillard}, {Orosz}, {Davis}, {Ebisawa}, \& {Schlegel}}]{Gou.2009}
{Gou}, L., {McClintock}, J.~E., {Liu}, J., {et~al.} 2009, \apj, 701, 1076,
  \dodoi{10.1088/0004-637X/701/2/1076}

\bibitem[{{Hawley} {et~al.}(1995){Hawley}, {Gammie}, \& {Balbus}}]{Hawley.1995}
{Hawley}, J.~F., {Gammie}, C.~F., \& {Balbus}, S.~A. 1995, \apj, 440, 742,
  \dodoi{10.1086/175311}

\bibitem[{{Janiuk} {et~al.}(2008){Janiuk}, {Proga}, \&
  {Kurosawa}}]{Janiuk.2008}
{Janiuk}, A., {Proga}, D., \& {Kurosawa}, R. 2008, \apj, 681, 58,
  \dodoi{10.1086/588375}

\bibitem[{{Kaaz} {et~al.}(2022){Kaaz}, {Murguia-Berthier}, {Chatterjee},
  {Liska}, \& {Tchekhovskoy}}]{Nicholas.2022}
{Kaaz}, N., {Murguia-Berthier}, A., {Chatterjee}, K., {Liska}, M., \&
  {Tchekhovskoy}, A. 2022, arXiv e-prints, arXiv:2201.11753.
\newblock \doarXiv{2201.11753}

\bibitem[{{Komissarov} \& {Barkov}(2009)}]{Komissarov.2009}
{Komissarov}, S.~S., \& {Barkov}, M.~V. 2009, \mnras, 397, 1153,
  \dodoi{10.1111/j.1365-2966.2009.14831.x}

\bibitem[{{Krolik} {et~al.}(2005){Krolik}, {Hawley}, \& {Hirose}}]{Krolik.2005}
{Krolik}, J.~H., {Hawley}, J.~F., \& {Hirose}, S. 2005, \apj, 622, 1008,
  \dodoi{10.1086/427932}

\bibitem[{{Lalakos} {et~al.}(2022){Lalakos}, {Gottlieb}, {Kaaz}, {Chatterjee},
  {Liska}, {Christie}, {Tchekhovskoy}, {Zhuravleva}, \&
  {Nokhrina}}]{Lalakos.2022}
{Lalakos}, A., {Gottlieb}, O., {Kaaz}, N., {et~al.} 2022, \apjl, 936, L5,
  \dodoi{10.3847/2041-8213/ac7bed}

\bibitem[{{Lee} \& {Ramirez-Ruiz}(2006)}]{Lee.2006}
{Lee}, W.~H., \& {Ramirez-Ruiz}, E. 2006, \apj, 641, 961,
  \dodoi{10.1086/500533}

\bibitem[{{Liu} {et~al.}(2008){Liu}, {McClintock}, {Narayan}, {Davis}, \&
  {Orosz}}]{Liu.2008}
{Liu}, J., {McClintock}, J.~E., {Narayan}, R., {Davis}, S.~W., \& {Orosz},
  J.~A. 2008, \apjl, 679, L37, \dodoi{10.1086/588840}

\bibitem[{{MacFadyen} \& {Woosley}(1999)}]{MacFadyen.1999}
{MacFadyen}, A.~I., \& {Woosley}, S.~E. 1999, \apj, 524, 262,
  \dodoi{10.1086/307790}

\bibitem[{Marrone {et~al.}(2006)Marrone, Moran, Zhao, \& Rao}]{Marrone.2006}
Marrone, D.~P., Moran, J.~M., Zhao, J.-H., \& Rao, R. 2006, The Astrophysical
  Journal, 654, L57, \dodoi{10.1086/510850}

\bibitem[{{McKinney} \& {Gammie}(2004)}]{McKinney.2004}
{McKinney}, J.~C., \& {Gammie}, C.~F. 2004, \apj, 611, 977,
  \dodoi{10.1086/422244}

\bibitem[{{McKinney} {et~al.}(2012){McKinney}, {Tchekhovskoy}, \&
  {Blandford}}]{McKinney.2012}
{McKinney}, J.~C., {Tchekhovskoy}, A., \& {Blandford}, R.~D. 2012, \mnras, 423,
  3083, \dodoi{10.1111/j.1365-2966.2012.21074.x}

\bibitem[{{Melia}(1992)}]{Melia.1992}
{Melia}, F. 1992, \apjl, 387, L25, \dodoi{10.1086/186297}

\bibitem[{{Mishra} {et~al.}(2020){Mishra}, {Begelman}, {Armitage}, \&
  {Simon}}]{Mishra.2020}
{Mishra}, B., {Begelman}, M.~C., {Armitage}, P.~J., \& {Simon}, J.~B. 2020,
  \mnras, 492, 1855, \dodoi{10.1093/mnras/stz3572}

\bibitem[{{Morris} \& {Serabyn}(1996)}]{Morris.1996}
{Morris}, M., \& {Serabyn}, E. 1996, \araa, 34, 645,
  \dodoi{10.1146/annurev.astro.34.1.645}

\bibitem[{{Murguia-Berthier} {et~al.}(2020){Murguia-Berthier}, {Batta},
  {Janiuk}, {Ramirez-Ruiz}, {Mandel}, {Noble}, \& {Everson}}]{Murguia.2020}
{Murguia-Berthier}, A., {Batta}, A., {Janiuk}, A., {et~al.} 2020, \apjl, 901,
  L24, \dodoi{10.3847/2041-8213/abb818}

\bibitem[{{Narayan} {et~al.}(2003){Narayan}, {Igumenshchev}, \&
  {Abramowicz}}]{Narayan.2003}
{Narayan}, R., {Igumenshchev}, I.~V., \& {Abramowicz}, M.~A. 2003, \pasj, 55,
  L69, \dodoi{10.1093/pasj/55.6.L69}

\bibitem[{{Narayan} {et~al.}(1998){Narayan}, {Mahadevan}, {Grindlay}, {Popham},
  \& {Gammie}}]{Narayan.1998}
{Narayan}, R., {Mahadevan}, R., {Grindlay}, J.~E., {Popham}, R.~G., \&
  {Gammie}, C. 1998, \apj, 492, 554, \dodoi{10.1086/305070}

\bibitem[{{Narayan} \& {Yi}(1994)}]{Narayan.1994.ADAF}
{Narayan}, R., \& {Yi}, I. 1994, \apjl, 428, L13, \dodoi{10.1086/187381}

\bibitem[{{Narayan} \& {Yi}(1995)}]{Narayan.1995.ADAF}
---. 1995, \apj, 452, 710, \dodoi{10.1086/176343}

\bibitem[{{Narayan} {et~al.}(1995){Narayan}, {Yi}, \&
  {Mahadevan}}]{Narayan.1995}
{Narayan}, R., {Yi}, I., \& {Mahadevan}, R. 1995, \nat, 374, 623,
  \dodoi{10.1038/374623a0}

\bibitem[{{Novikov} \& {Thorne}(1973)}]{Novikov.1973.NT}
{Novikov}, I.~D., \& {Thorne}, K.~S. 1973, in Black Holes (Les Astres Occlus),
  343--450

\bibitem[{{Porth} {et~al.}(2017){Porth}, {Olivares}, {Mizuno}, {Younsi},
  {Rezzolla}, {Moscibrodzka}, {Falcke}, \& {Kramer}}]{Porth.2017}
{Porth}, O., {Olivares}, H., {Mizuno}, Y., {et~al.} 2017, Computational
  Astrophysics and Cosmology, 4, 1, \dodoi{10.1186/s40668-017-0020-2}

\bibitem[{{Porth} {et~al.}(2019){Porth}, {Chatterjee}, {Narayan}, {Gammie},
  {Mizuno}, {Anninos}, {Baker}, {Bugli}, {Chan}, {Davelaar}, {Del Zanna},
  {Etienne}, {Fragile}, {Kelly}, {Liska}, {Markoff}, {McKinney}, {Mishra},
  {Noble}, {Olivares}, {Prather}, {Rezzolla}, {Ryan}, {Stone}, {Tomei},
  {White}, {Younsi}, {Akiyama}, {Alberdi}, {Alef}, {Asada}, {Azulay}, {Baczko},
  {Ball}, {Balokovi{\'c}}, {Barrett}, {Bintley}, {Blackburn}, {Boland},
  {Bouman}, {Bower}, {Bremer}, {Brinkerink}, {Brissenden}, {Britzen},
  {Broderick}, {Broguiere}, {Bronzwaer}, {Byun}, {Carlstrom}, {Chael},
  {Chatterjee}, {Chen}, {Chen}, {Cho}, {Christian}, {Conway}, {Cordes},
  {Geoffrey}, {Crew}, {Cui}, {De Laurentis}, {Deane}, {Dempsey}, {Desvignes},
  {Doeleman}, {Eatough}, {Falcke}, {Fish}, {Fomalont}, {Fraga-Encinas},
  {Freeman}, {Friberg}, {Fromm}, {G{\'o}mez}, {Galison}, {Garc{\'\i}a},
  {Gentaz}, {Georgiev}, {Goddi}, {Gold}, {Gu}, {Gurwell}, {Hada}, {Hecht},
  {Hesper}, {Ho}, {Ho}, {Honma}, {Huang}, {Huang}, {Hughes}, {Ikeda}, {Inoue},
  {Issaoun}, {James}, {Jannuzi}, {Janssen}, {Jeter}, {Jiang}, {Johnson},
  {Jorstad}, {Jung}, {Karami}, {Karuppusamy}, {Kawashima}, {Keating},
  {Kettenis}, {Kim}, {Kim}, {Kim}, {Kino}, {Koay}, {Patrick}, {Koch}, {Koyama},
  {Kramer}, {Kramer}, {Krichbaum}, {Kuo}, {Lauer}, {Lee}, {Li}, {Li},
  {Lindqvist}, {Liu}, {Liuzzo}, {Lo}, {Lobanov}, {Loinard}, {Lonsdale}, {Lu},
  {MacDonald}, {Mao}, {Marrone}, {Marscher}, {Mart{\'\i}-Vidal}, {Matsushita},
  {Matthews}, {Medeiros}, {Menten}, {Mizuno}, {Moran}, {Moriyama},
  {Moscibrodzka}, {M{\"u}ller}, {Nagai}, {Nagar}, {Nakamura}, {Narayanan},
  {Natarajan}, {Neri}, {Ni}, {Noutsos}, {Okino}, {Oyama}, {{\"O}zel},
  {Palumbo}, {Patel}, {Pen}, {Pesce}, {Pi{\'e}tu}, {Plambeck}, {PopStefanija},
  {Preciado-L{\'o}pez}, {Psaltis}, {Pu}, {Ramakrishnan}, {Rao}, {Rawlings},
  {Raymond}, {Ripperda}, {Roelofs}, {Rogers}, {Ros}, {Rose}, {Roshanineshat},
  {Rottmann}, {Roy}, {Ruszczyk}, {Rygl}, {S{\'a}nchez},
  {S{\'a}nchez-Arguelles}, {Sasada}, {Savolainen}, {Schloerb}, {Schuster},
  {Shao}, {Shen}, {Small}, {Sohn}, {SooHoo}, {Tazaki}, {Tiede}, {Tilanus},
  {Titus}, {Toma}, {Torne}, {Trent}, {Trippe}, {Tsuda}, {van Bemmel}, {van
  Langevelde}, {van Rossum}, {Wagner}, {Wardle}, {Weintroub}, {Wex}, {Wharton},
  {Wielgus}, {Wong}, {Wu}, {Young}, {Young}, {Yuan}, {Yuan}, {Zensus}, {Zhao},
  {Zhao}, {Zhu}, \& {Event Horizon Telescope Collaboration}}]{Porth.2019}
{Porth}, O., {Chatterjee}, K., {Narayan}, R., {et~al.} 2019, \apjs, 243, 26,
  \dodoi{10.3847/1538-4365/ab29fd}

\bibitem[{{Proga} \& {Begelman}(2003{\natexlab{a}})}]{Proga.2003}
{Proga}, D., \& {Begelman}, M.~C. 2003{\natexlab{a}}, \apj, 582, 69,
  \dodoi{10.1086/344537}

\bibitem[{{Proga} \& {Begelman}(2003{\natexlab{b}})}]{Proga.2003b}
---. 2003{\natexlab{b}}, \apj, 592, 767, \dodoi{10.1086/375773}

\bibitem[{{Quataert}(2004)}]{Quataert.2004}
{Quataert}, E. 2004, \apj, 613, 322, \dodoi{10.1086/422973}

\bibitem[{{Rees}(1988)}]{Rees.1988}
{Rees}, M.~J. 1988, \nat, 333, 523, \dodoi{10.1038/333523a0}

\bibitem[{{Ressler} {et~al.}(2018){Ressler}, {Quataert}, \&
  {Stone}}]{Ressler.2018}
{Ressler}, S.~M., {Quataert}, E., \& {Stone}, J.~M. 2018, \mnras, 478, 3544,
  \dodoi{10.1093/mnras/sty1146}

\bibitem[{{Ressler} {et~al.}(2021){Ressler}, {Quataert}, {White}, \&
  {Blaes}}]{Ressler.2021}
{Ressler}, S.~M., {Quataert}, E., {White}, C.~J., \& {Blaes}, O. 2021, \mnras,
  504, 6076, \dodoi{10.1093/mnras/stab311}

\bibitem[{{Ressler} {et~al.}(2020){Ressler}, {White}, {Quataert}, \&
  {Stone}}]{Ressler.2020}
{Ressler}, S.~M., {White}, C.~J., {Quataert}, E., \& {Stone}, J.~M. 2020,
  \apjl, 896, L6, \dodoi{10.3847/2041-8213/ab9532}

\bibitem[{{Shakura} \& {Sunyaev}(1973)}]{Shakura.1973.standard.disk}
{Shakura}, N.~I., \& {Sunyaev}, R.~A. 1973, \aap, 500, 33

\bibitem[{{Shapiro} \& {Lightman}(1976)}]{Shapiro.1976}
{Shapiro}, S.~L., \& {Lightman}, A.~P. 1976, \apj, 204, 555,
  \dodoi{10.1086/154203}

\bibitem[{{Smith} {et~al.}(2002){Smith}, {Heindl}, \& {Swank}}]{Smith.2002}
{Smith}, D.~M., {Heindl}, W.~A., \& {Swank}, J.~H. 2002, \apj, 569, 362,
  \dodoi{10.1086/339167}

\bibitem[{Stanzione {et~al.}(2020)Stanzione, West, Evans, Minyard, Ghattas, \&
  Panda}]{stanzione2020frontera}
Stanzione, D., West, J., Evans, R.~T., {et~al.} 2020, in Practice and
  Experience in Advanced Research Computing, 106--111,
  \dodoi{10.1145/3311790.3396656}

\bibitem[{{Stirling} {et~al.}(2001){Stirling}, {Spencer}, {de la Force},
  {Garrett}, {Fender}, \& {Ogley}}]{Stirling.2001}
{Stirling}, A.~M., {Spencer}, R.~E., {de la Force}, C.~J., {et~al.} 2001,
  \mnras, 327, 1273, \dodoi{10.1046/j.1365-8711.2001.04821.x}

\bibitem[{{Sukov{\'a}} {et~al.}(2017){Sukov{\'a}}, {Charzy{\'n}ski}, \&
  {Janiuk}}]{Sukova.2017}
{Sukov{\'a}}, P., {Charzy{\'n}ski}, S., \& {Janiuk}, A. 2017, \mnras, 472,
  4327, \dodoi{10.1093/mnras/stx2254}

\bibitem[{{Sukov{\'a}} \& {Janiuk}(2015)}]{Sukova.2015}
{Sukov{\'a}}, P., \& {Janiuk}, A. 2015, \mnras, 447, 1565,
  \dodoi{10.1093/mnras/stu2544}

\bibitem[{{Tauris} \& {van den Heuvel}(2006)}]{Tauris.2006}
{Tauris}, T.~M., \& {van den Heuvel}, E.~P.~J. 2006, {Formation and evolution
  of compact stellar X-ray sources}, Vol.~39, 623--665

\bibitem[{{Tchekhovskoy}(2015)}]{Tchekhovskoy.2015}
{Tchekhovskoy}, A. 2015, in Astrophysics and Space Science Library, Vol. 414,
  The Formation and Disruption of Black Hole Jets, ed. I.~{Contopoulos},
  D.~{Gabuzda}, \& N.~{Kylafis}, 45, \dodoi{10.1007/978-3-319-10356-3_3}

\bibitem[{{Tchekhovskoy} {et~al.}(2008){Tchekhovskoy}, {McKinney}, \&
  {Narayan}}]{Tchekhovskoy.2008}
{Tchekhovskoy}, A., {McKinney}, J.~C., \& {Narayan}, R. 2008, \mnras, 388, 551,
  \dodoi{10.1111/j.1365-2966.2008.13425.x}

\bibitem[{{Tchekhovskoy} {et~al.}(2012){Tchekhovskoy}, {McKinney}, \&
  {Narayan}}]{Tchekhovskoy.2012a}
{Tchekhovskoy}, A., {McKinney}, J.~C., \& {Narayan}, R. 2012, in Journal of
  Physics Conference Series, Vol. 372, Journal of Physics Conference Series,
  012040, \dodoi{10.1088/1742-6596/372/1/012040}

\bibitem[{{Tchekhovskoy} {et~al.}(2014){Tchekhovskoy}, {Metzger}, {Giannios},
  \& {Kelley}}]{Tchekhovskoy14}
{Tchekhovskoy}, A., {Metzger}, B.~D., {Giannios}, D., \& {Kelley}, L.~Z. 2014,
  \mnras, 437, 2744, \dodoi{10.1093/mnras/stt2085}

\bibitem[{{Tchekhovskoy} {et~al.}(2011){Tchekhovskoy}, {Narayan}, \&
  {McKinney}}]{Tchekhovskoy.2011}
{Tchekhovskoy}, A., {Narayan}, R., \& {McKinney}, J.~C. 2011, \mnras, 418, L79,
  \dodoi{10.1111/j.1745-3933.2011.01147.x}

\bibitem[{{The Event Horizon Telescope Collaboration et al.}(2022)}]{EHT.2022}
{The Event Horizon Telescope Collaboration et al.} 2022, \apjl, 930, 21,
  \dodoi{10.3847/2041-8213/ac6674}

\bibitem[{{White} {et~al.}(2016){White}, {Stone}, \& {Gammie}}]{White.2016}
{White}, C.~J., {Stone}, J.~M., \& {Gammie}, C.~F. 2016, \apjs, 225, 22,
  \dodoi{10.3847/0067-0049/225/2/22}

\bibitem[{{Woosley}(1993)}]{Woosley.1993}
{Woosley}, S.~E. 1993, \apj, 405, 273, \dodoi{10.1086/172359}

\bibitem[{{Yuan} \& {Narayan}(2014)}]{Yuan.2014}
{Yuan}, F., \& {Narayan}, R. 2014, \araa, 52, 529,
  \dodoi{10.1146/annurev-astro-082812-141003}

\bibitem[{{Zauderer} {et~al.}(2013){Zauderer}, {Berger}, {Margutti}, {Pooley},
  {Sari}, {Soderberg}, {Brunthaler}, \& {Bietenholz}}]{Zauderer13}
{Zauderer}, B.~A., {Berger}, E., {Margutti}, R., {et~al.} 2013, \apj, 767, 152,
  \dodoi{10.1088/0004-637X/767/2/152}

\end{thebibliography}
\bibliographystyle{aasjournal}



\end{document}